\documentclass[conference]{IEEEtran}
\IEEEoverridecommandlockouts
\usepackage{cite}

\usepackage{amsmath,amsfonts,amsthm}

\usepackage{algorithmic}
\usepackage{graphicx}
\usepackage{textcomp}
\usepackage{cite}
\usepackage{amsmath,amssymb,amsfonts}
\usepackage{algorithmic}
\usepackage{graphicx}
\usepackage{mathtools}
\usepackage{textcomp}
\usepackage{bbold}
\usepackage{enumerate}
\usepackage{xcolor}
\usepackage{bm}
\usepackage{algorithm}
\usepackage{babel,blindtext}
\usepackage{subcaption}
\usepackage{tabularx,booktabs}

\newtheorem{theorem}{Theorem}
\newtheorem{lemma}{Lemma}

\newtheorem{assumption}{Assumption}

\newtheorem{definition}{Definition}
\newtheorem{remark}{Remark}

\begin{document}

\title{Network Slicing: Market Mechanism and Competitive Equilibria
{\footnotesize }
\thanks{This paper appeared in INFOCOM 2023.}
\thanks{The research work was supported by the Office of Naval Research under project numbers N00014-19-1-2566, N00173-21-1-G006 and by the National Science Foundation under the project number CNS-2128530.}}

\author{\IEEEauthorblockN{Panagiotis Promponas, and
Leandros Tassiulas
 \\
 \IEEEauthorblockA{Department of Electrical Engineering and Institute for Network Science, Yale University, USA}
 \{panagiotis.promponas, leandros.tassiulas\}@yale.edu}
}

\maketitle

\begin{abstract}
Towards addressing spectral scarcity and enhancing resource utilization in 5G networks, network slicing is a promising technology to establish end-to-end virtual networks without requiring additional infrastructure investments. By leveraging Software Defined Networks (SDN) and Network Function Virtualization (NFV), we can realize slices completely isolated and dedicated to satisfy the users' diverse Quality of Service (QoS) prerequisites and Service Level Agreements (SLAs). This paper focuses on the technical and economic challenges that emerge from the application of the network slicing architecture to real-world scenarios. We consider a market where multiple Network Providers (NPs) own the physical infrastructure and offer their resources to multiple Service Providers (SPs). Then, the SPs offer those resources as slices to their associated users. We propose a holistic iterative model for the network slicing market along with a clock auction that converges to a robust $\epsilon$-competitive equilibrium. At the end of each cycle of the market, the slices are reconfigured and the SPs aim to learn the private parameters of their users. Numerical results are provided that validate and evaluate the convergence of the clock auction and the capability of the proposed market architecture to express the incentives of the different entities of the system.
\end{abstract}

\vspace{0.2cm}

\begin{IEEEkeywords}
Network Slicing, Mechanism Design, Network Economics, Bayesian Inference
\end{IEEEkeywords}

\section{Introduction}
The ascending trend of the volume of the data traffic, as well as the vast number of connected devices, puts pressure on the industries to enhance resource utilization in 5G wireless networks. With the advent of 5G networks and Internet of Things (IoT), researchers aim at a technological transformation to simultaneously improve throughput, extend network coverage and augment the users' quality of service without wasting valuable resources. Despite the significant advances brought by the enhanced network architectures and technologies, spectral scarcity will still impede the realization of the full potential of 5G technology.
\par
In the future 5G networks, verticals need distinct network services as they may differ in their Quality of Service (QoS) requirements, Service Level Agreements (SLAs), and key performance indicators (KPIs). Such a need highlights the inefficiency of the previous architecture technologies which were based on a "one network fits all" nature. In this direction, network slicing is a promising technology that enables the transition from one-size-fits-all to one-size-per-service abstraction \cite{zhang2019adaptive}, which is customized for the distinct use cases in a contemporary 5G network model.
\par
Using Software Defined Networks (SDN) and Network Function Virtualization (NFV), those slices are associated with completely isolated resources that can be tailored on-demand to satisfy the diverse QoS prerequisites and SLAs. Resource allocation in network slicing plays a pivotal role in load balancing, resource utilization and networking performance \cite{su2019resource}. Nevertheless, such a resource allocation model faces various challenges in terms of isolation, customization, and end-to-end coordination which involves both the core but also the Radio Access Network (RAN) \cite{qin2020network}.  
\par
In a typical network slicing scenario, multiple Network Providers (NPs), own the physical infrastructure and offer their resources to multiple Service Providers (SPs). Possible services of the SPs include e-commerce, video, gaming, virtual reality, wearable smart devices, and other IoT devices. The SPs offer their resources as completely isolated slices to their associated users. Thereby, such a system contains three types of actors that interact with each other and compete for the same resources, either monetary or networking. This paper focuses on the technical and economic challenges that emerge from the application of this architecture to real-world scenarios.  


\subsection{Related Work}

\par 




\par 
\textbf{User Satisfaction \& Sigmoid Functions:} 
Network applications can be separated into elastic (e.g. email, text file transfer) and inelastic (e.g. audio/video phone, video conference, tele-medicine) \cite{pham2016network}. Utilities for elastic applications are modeled as concave functions that increase with the resources with diminishing returns \cite{pham2016network}. On the other hand, the utility function for an inelastic traffic is modeled as a non-concave and usually as a sigmoid function. Such non-concavities impose challenges for the optimization of a network, but are suitable with the 5G era where the services may differ in their QoS requirements \cite{lee2005non}. In that direction, multiple works in the literature employ sigmoid utility functions for the network users \cite{liu2013theoretical, lieto2020strategic, zhu2008nonlinear,gao2016virtualization, lee2014qoe, hemmati2017qoe, lee2005non, tan2015utility, papavassiliou2020paradigm}. Nevertheless, all of these works consider either one SP and model the interaction between the users, or multiple SPs that compete for a fixed amount of resources (e.g. bandwidth).

\textbf{Network Slicing in 5G Networks:}
Network slicing introduces various challenges to the resource allocation in 5G networks in terms of isolation, customization, elasticity, and end-to-end coordination \cite{su2019resource}. Most surveys on network slicing investigate its multiple business models motivated by 5G, the fundamental architecture of a slice and the state-of-the-art algorithms of network slicing \cite{su2019resource, afolabi2018network, vassilaras2017algorithmic}. Microeconomic theories such as non-cooperative games and/or mechanism design arise as perfect tools to model the trading of network infrastructure and radio resources that takes place in network slicing \cite{habiba2018auction, zhang2016double, gao2016virtualization, fu2010wireless}.

\textbf{Mechanism Design in Network Slicing:}
 Multiple auction mechanisms have been used to identify the business model of a network slicing market 
(see a survey in \cite{habiba2018auction}). Contrary to our work, the majority of the literature considers a single-sided auction, a model that assumes that a single NP owns the whole infrastructure of the market \cite{gao2016virtualization, fu2010wireless, ahmadi2016virtualization,cao2015power,zhu2015virtualization,zhu2017wireless}. For example, \cite{gao2016virtualization} considers a Vickrey–Clarke–Groves (VCG) auction-based model where the NP plays the role of an auctioneer and distributes discrete physical resource blocks.
We find  \cite{qin2020network} and \cite{zhang2016double} to be closer to our work, since the authors employ the double-sided auction introduced by \cite{iosifidis2014double} to maximize the social welfare of a system with multiple NPs. Contrary to our work, the auction proposed in \cite{iosifidis2014double} assumes concave utility functions for the different actors and requires the computation of their gradients for its convergence. The aforementioned assumptions might lead to an over-simplification of a more complex networking architecture (e.g. that of the network slicing model) where the utility function for a user with inelastic traffic is expressed as a sigmoid function \cite{gao2016virtualization} and that of an SP as an optimization problem \cite{qin2020network}.

\subsection{Contributions}

Our work develops an iterative market model for the network slicing architecture, where multiple NPs with heterogeneous Radio Access Technologies (RATs), own the physical infrastructure and offer their resources to multiple SPs. The latter offer the resources as slices to their associated users. Specifically, we propose a $five$-step iterative model for the network slicing market that converges to a robust $\epsilon$-competitive equilibrium even when the utility functions of the different actors are non-concave. In every cycle of the proposed model, the slices are reconfigured and the SPs learn the private parameters of their associated end-users to make the equilibrium of the next cycle more efficient. The introduced market model, can be seen as a framework that suits well to various networking problems where $three$ types of actors are involved: those who own the physical infrastructure, those who lease part of it to sell services and those who enjoy the services (e.g. data-offloading \cite{iosifidis2014double}).

For the interaction between the SPs and the NPs and for the convergence of the market to an equilibrium, we propose an iterative clock auction. Such dynamic auctions are used in the literature to auction divisible goods \cite{ausubel2004auctioning, ausubel2006clock}. The key differentiating aspects of the proposed auction, are (i) the relaxation of the common assumptions that the utility functions are concave and their gradients can be analytically computed, (ii) it provides highly usable price discovery, and (iii) it is a double-sided auction, and thus appropriate for a market with multiple NPs. Numerical results are provided that validate and evaluate the convergence of the clock auction and the capability of the proposed market architecture to express the incentives of the different entities of the system.

\section{Market Model \& Incentives} \label{Section II}

In this section we describe the different entities of the network slicing market and their conflicting incentives.
\vspace{-11pt}

\subsection{Market Model}
A typical slicing system model \cite{qin2020network, su2019resource, afolabi2018network, vassilaras2017algorithmic} consists of multiple SPs represented by $\mathcal{M} = \{1,2,\dots,M\}$ and multiple NPs that own RANs of possibly different RATs, represented by a set $\mathcal{K} = \{ 1,2,\dots,K \}$. Each SP owns a slice with a predetermined amount of isolated resources (e. g., bandwidth) and is associated with a set of users, $\mathcal{U}_m$, that serves through its slices. For the rest of the paper and without loss of generality we assume that each NP owns exactly one RAN, so we use the terms RAN and NP interchangeably.

\subsubsection{Network Providers}
    The multiple NPs of the system can quantify their radio resources as the performance level of the same network metric (e.g., downlink throughput) \cite{qin2020network}. Let $x_{(m,k)}$ denote the amount of resources NP $k$ allocates to SP $m$, and the vector $\bm{x}_m := (x_{(m,k)})_{k \in \mathcal{K}}$ to denote the amount of resources $m$ gets from every NP. Without loss of generality \cite{qin2020network}, capacity $C_k$ limits the amount of resources that can be offered from NP $k$, i.e., $\sum_{m=1}^{M} x_{(m,k)} \le C_k$. Let $\bm{C} = (C_k)_{k \in \mathcal{K}}$. For the rest of the paper, we assume that there is a constant cost related to operation and management overheads induced to the NP. The main goal of every NP $k$ is to maximize its profits by adjusting the price per unit of resources, denoted by $c_k$.

\subsubsection{Service Providers \& Associated Users}
The main goal of an SP is to purchase resources from a single or multiple NPs in order to maximize its profit, which depends on its associated users' satisfaction. The connectivity of a user $i \in \mathcal{U}_m$ is denoted by a vector $\bm{\beta}_i = (\beta_{(k,i)})_{k\in \mathcal{K}}$, where $\beta_{(k,i)}$ is a non-negative number representing factors such as the link quality i.e., numbers in $(0, 1]$ that depend on the path loss. Moreover, each user $i$ of the SP $m$, is associated with a service class, $\mathbb{c}(i)$, depending on their preferences. We denote the set of the possible service classes of SP $m$ as $\mathcal{C}^{m} = \{C^m_1, \dots, C^m_{c_m}\}$ and thus $\mathbb{c}(i) \in \mathcal{C}^{m}, \quad \forall i \in \mathcal{U}_m$. Each SP $m$, is trying to distribute the resources purchased from the NPs, i.e., $\bm{x}_m$, to maximize its profit. This process, referred to as intra-slice resource allocation, is described in detail in Section \ref{sec:intraslice}.

Throughout the paper, we assume that the number of users of every SP $m$, i.e., $|\mathcal{U}_m|$, is much greater than the number of SPs, which is much greater than the number of NPs in the market. 
This assumption is made often in the mechanism design literature and is sufficient to ensure that the end-users and the SPs have limited information of the market \cite{shen2018first, iosifidis2014double}. The latter let us consider them as price-takers.
In the following section, we describe in detail the intra-slice resource allocation problem from the perspective of an SP who tries to maximize the satisfaction of its associated users.

\subsection{Intra-Slice Resource Allocation}
\label{sec:intraslice}

The problem of the intra-slice resource allocation concerns the distribution of the resources, $\bm{x}_m$, from the SP $m$ to its associated users. Specifically, every SP $m$ allocates a portion of $x_{(m,k)}$ to its associated user $i$, denoted as $r_{(k,i)}$. Let $\bm{r}_i := (r_{(k,i)})_{k \in \mathcal{K}}$ and $\bm{r}_m := (\bm{r}_i)_{  i \in \mathcal{U}_m}$. For ease of notation, the resources, $\bm{r}_i$, of a user $i \in \mathcal{U}_m$, as well as the connectivities, $\bm{\beta}_i$, are not indexed by $m$ because $i$ is assumed to be a unique identifier for the user. Although every user $i$ is assigned with ${r}_{(k,i)}$ resources from RAN $k$, because of its connectivity $\bm{\beta}_i$, the aggregated amount of resources it gets is $z_i := \bm{\beta_i^T}\bm{r}_i$. Moreover, let $\bm{z}_m := ({z_i})_{  i \in \mathcal{U}_m}$. In a feasible intra-slice allocation it should hold that $\bm{x}_m  \succeq \sum_{i \in \mathcal{U}_m} \bm{r}_i $ for each SP $m$.

Every SP should distribute the obtained resources among its users to maximize their satisfaction. Towards providing intuition behind the employment of sigmoidal functions in the literature to model user satisfaction (e.g. see \cite{liu2013theoretical, lieto2020strategic, zhu2008nonlinear,gao2016virtualization, lee2014qoe, hemmati2017qoe, lee2005non, tan2015utility}), note that by making the same assumption as logistic regression, we model the logit\footnote{The logit function is defined as $logit(p) = log(\frac{p}{1-p})$.} of the probability that a user is satisfied, as a linear function of the resources. Hence, the probability that user $i$ is satisfied with the amount of resources $z_i$, say $P[QoS\_sat_i]$, satisfies 
$log(\frac{P[QoS\_sat_i]}{1 - P[QoS\_sat_i]}) = t^{z}_{\mathbb{c}(i)}(z_i - k_{\mathbb{c}(i)})$ and thus:
\begin{equation}
P[QoS\_sat_i] =  \frac{e^{t^{z}_{\mathbb{c}(i)}(z_i-k_{\mathbb{c}(i)})}}{1+e^{t^{z}_{\mathbb{c}(i)}(z_i-k_{\mathbb{c}(i)})}},    
\end{equation}
where $k_{\mathbb{c}(i)} \ge 0$ denotes the prerequisite amount of resources of the user $i$ and  $t^{z}_{\mathbb{c}(i)} \ge 0$ expresses how "tight" this prerequisite is. Note that the probability of a user being satisfied with respect to the value of $z_i$, is a sigmoid function with inflection point $k_{\mathbb{c}(i)}$. We assume that the user's service class fully determines its private parameters, hence every user $i \in \mathbb{c}(i)$ has QoS prerequisite $k_{\mathbb{c}(i)}$ and sensitivity parameter $t^{z}_{\mathbb{c}(i)}$. These parameters are unknown to the users, so the SP's goal to eventually learn them is challenging (Section \ref{sec:learning}).




Given the previous analysis, the aggregated satisfaction of the users of the SP $m$ is $u_m(\bm{r}_m) := \sum_{i \in \mathcal{U}_m }   u_{i}(\bm{r}_i) $ (\cite{lee2014qoe}, \cite{lieto2020strategic}), where 
\begin{equation}
u_{i}(\bm{r}_i) := \frac{e^{t^{z}_{\mathbb{c}(i)} (\bm{\beta}_i^T \bm{r}_i-k_{\mathbb{c}(i)})}}{1+e^{t^{z}_{\mathbb{c}(i)} (\bm{\beta}_i^T \bm{r}_i-k_{\mathbb{c}(i)})}}  .  
\end{equation}
Note that the function $u_{i}(\cdot)$ can be expressed as a function of $z_i$ as well. With a slight abuse of notation, we switch between the two by changing the input variable. We can write the final optimization problem for the intra-slice allocation of SP $m$ as: 
\begin{equation*}
\begin{aligned} 
\textbf{(IN-SL): $\quad$} \max_{\bm{r}_m} \quad &   u_m(\bm{r}_m) \\
\textrm{s.t.} \quad & \bm{r}_i \succeq \bm{0}, \quad \forall i \in \mathcal{U}_m\\
  \quad & \bm{x}_m \succeq \sum_{i \in \mathcal{U}_m} \bm{r}_i \\
\end{aligned}
\normalsize   
\end{equation*}
In case the amount of resources obtained from every NP, $\bm{x_m}$, is not given, SP $m$ can optimize it together with the intra-slice resource allocation. 
Hence, SP $m$ can solve the following problem
\begin{equation*}
 \begin{aligned} 
\textbf{(P): $\quad$} \max_{\bm{r}_m, \bm{x_m}} \quad &  {\Psi}_m(\bm{r}_m, \bm{x_m}) := u_m(\bm{r}_m) - \bm{c}^T \bm{x}_{m}\\
\textrm{s.t.} \quad & \bm{r}_i \succeq \bm{0}, \quad \forall i \in \mathcal{U}_m\\
  \quad & \bm{x}_m \succeq \sum_{i \in \mathcal{U}_m} \bm{r}_i \\
\end{aligned}   
\end{equation*}
Recall that $c_k$ denotes the price per unit of resources announced from every NP $k$. In Problem $\bm{P}$, the objective function $\Psi_m$ can be thought of as the profit of SP $m$.
Let the solution of the above problem be $\psi_m^{*}$.

Problems \textbf{IN-SL} and \textbf{P} are maximization problems of a summation of sigmoid functions over a linear set of constraints. In \cite{udell2013maximizing} the problem of maximizing a sum of sigmoid functions over a convex constraint set is addressed. This work shows that this problem is generally NP-hard and it proposes an approximation algorithm, using a branch-and-bound method, to find an approximate solution to the sigmoid programming problem. 

In the rest of the section, we study three variations of problem \textbf{P}. Specifically, in Section \ref{sec:pricingmechanism}, we study the case where the end-users are charged to get the resources from the SPs and in Sections \ref{sec:regularization} and \ref{sec:concavification} we regularize and concavify \textbf{P} respectively, something that will facilitate the analysis of the rest of the paper.  

\subsubsection{Price Mechanism in \textbf{P}} 
\label{sec:pricingmechanism}

In this subsection we argue that Problem \textbf{P} is expressive enough to capture the case where every user $i$ is charged for its assigned resources. Let $p_i$ be the amount of money that user $i$ should pay to receive the $z_i$ resources. In that case, the SPs should modify Problems \textbf{IN-SL} and \textbf{P} accordingly. First, note that user $i$'s satisfaction may depend also on $p_i$. Similarly with the previous section, we can express the satisfaction of user $i$ with respect to the price $p_i$ using a sigmoid function as
$
    P[price\_sat_i] =  \frac{1}{1+e^{t^{p}_{\mathbb{c}(i)}(p_i-b_{\mathbb{c}(i)})}}, 
$
where $b_{\mathbb{c}(i)} \ge 0$ is the budget of the user $i$ for the prerequisite resources $k_{\mathbb{c}(i)}$, and  $t^{p}_{\mathbb{c}(i)}  \ge 0$ expresses how "tight" is this budget. We can now model the \textit{acceptance probability} function \cite{lieto2020strategic} as
$
    P[sat_i]  = P[price\_sat_i] P[QoS\_sat_i] ,
$
and hence the expected total revenue, or the new utility of SP $m$, $u^{'}_m$, is modeled as
\begin{equation}
    \label{tt2} u^{'}_m(\bm{r}_m, \bm{p}_m) :=
     \sum_{i \in \mathcal{U}_m } P[sat_i] p_i.
\end{equation}
From Eq. \eqref{tt2}, it is possible for SP $m$ to immediately determine the optimal price $\hat{p}_i$ to ask from any user $i \in \mathcal{U}_m$. This follows from the fact that for positive $p_i$ the function admits a unique critical point, $\hat{p}$. 
Therefore, by just adding proper coefficients to the terms of Problem \textbf{IN-SL} and \textbf{P}, we can embed a pricing mechanism for the end-users in the model. For the rest of the paper, without loss of generality in our model, we assume that the end-users are not charged for the obtained resources.




\subsubsection{Regularization of $\bm{P}$}
\label{sec:regularization}

\par
We can regularize Problem $\bm{P}$, with a small positive $\lambda_m$. In that manner, we encourage dense solutions and hence we avoid situations where a problem in one RAN completely disrupts the operation of the SP. 
\begin{equation*}
\begin{aligned} 
\textbf{($\bm{\bar{P})}$: $\quad$} \max_{\bm{r}_m, \bm{x_m}} \quad &   {\Psi}_m(\bm{r}_m, \bm{x_m}) - \lambda_m \|\bm{x}_m\|_2^2\\
\textrm{s.t.} \quad & \bm{r}_i \succeq \bm{0}, \quad \forall i \in \mathcal{U}_m\\
  \quad & \bm{x}_m \succeq \sum_{i \in \mathcal{U}_m} \bm{r}_i \\
\end{aligned}    
\end{equation*}
In the regularized problem $\bm{\bar{P}}$, note that larger values of $\lambda_m$ penalize the vectors $\bm{x}_m$ with greater L2 norms. Let the solution of Problem \textbf{$\bm{\bar{P}}$} be $\bar{\psi}_m^{*}$. The Lemma below, shows that for small $\lambda_m$, the optimal values $\bar{\psi}_m^{*}$ and ${\psi}_m^{*}$ are close. Its proof is simple and thus ommited for brevity.

\begin{lemma}
\label{lemma_regularization_optimality}
Let $(\bm{r}_m^*, \bm{x}_m^*)$ and $(\bm{\bar{r}}_m^*, \bm{\bar{x}}_m^*)$ be solutions of Problems $\bm{P}$ and $\bm{\bar{P}}$ respectively. Then, 
\[ {\psi^*_m} - \lambda_m \|{{x}}_m^*\|_2^2 \le \bar{\psi}_m^* \le \psi_m^*  -  \lambda_m\|\bar{x}_m^*\|_2^2  \]
\end{lemma}

Lemma \ref{lemma_regularization_optimality}, proves that the regularization of $\bm{P}$ was (almost) without loss of optimality. In the next section, we proceed by concavifying Problem $\bm{\bar{P}}$. The new concavified problem will be a fundamental building block of the auction analysis in Section \ref{sec:auction}.

\subsubsection{Concavification of $\bar{\bm{P}}$}
\label{sec:concavification}

\par

\begin{figure}[t]
\begin{subfigure}[b]{0.32\linewidth}
\includegraphics[height=4.5cm,width=\linewidth]{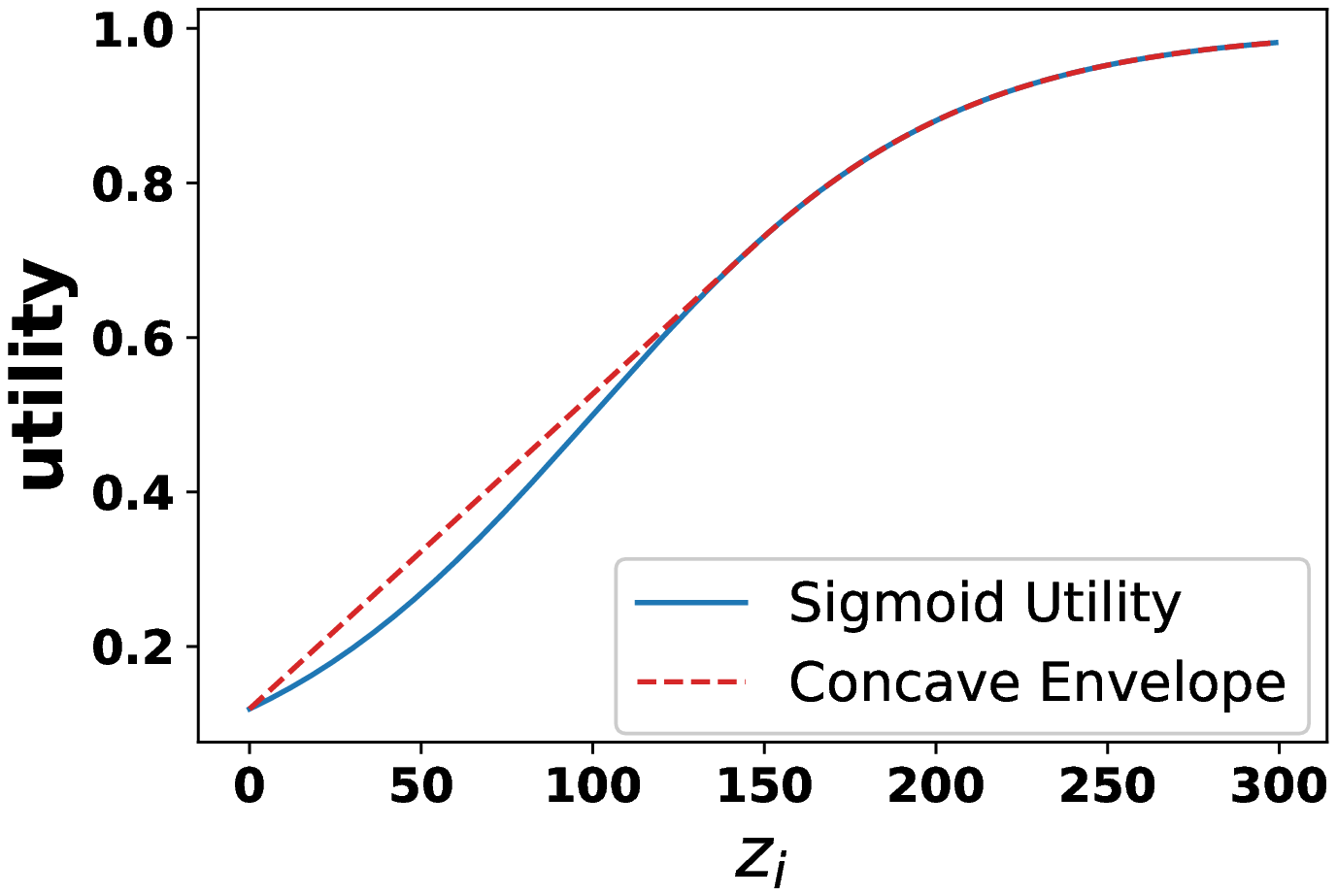}
\vspace{-20pt}
\caption{}
\label{fig:0a}
\end{subfigure}
\begin{subfigure}[b]{0.32\linewidth}
\includegraphics[height=4.5cm,width=\linewidth]{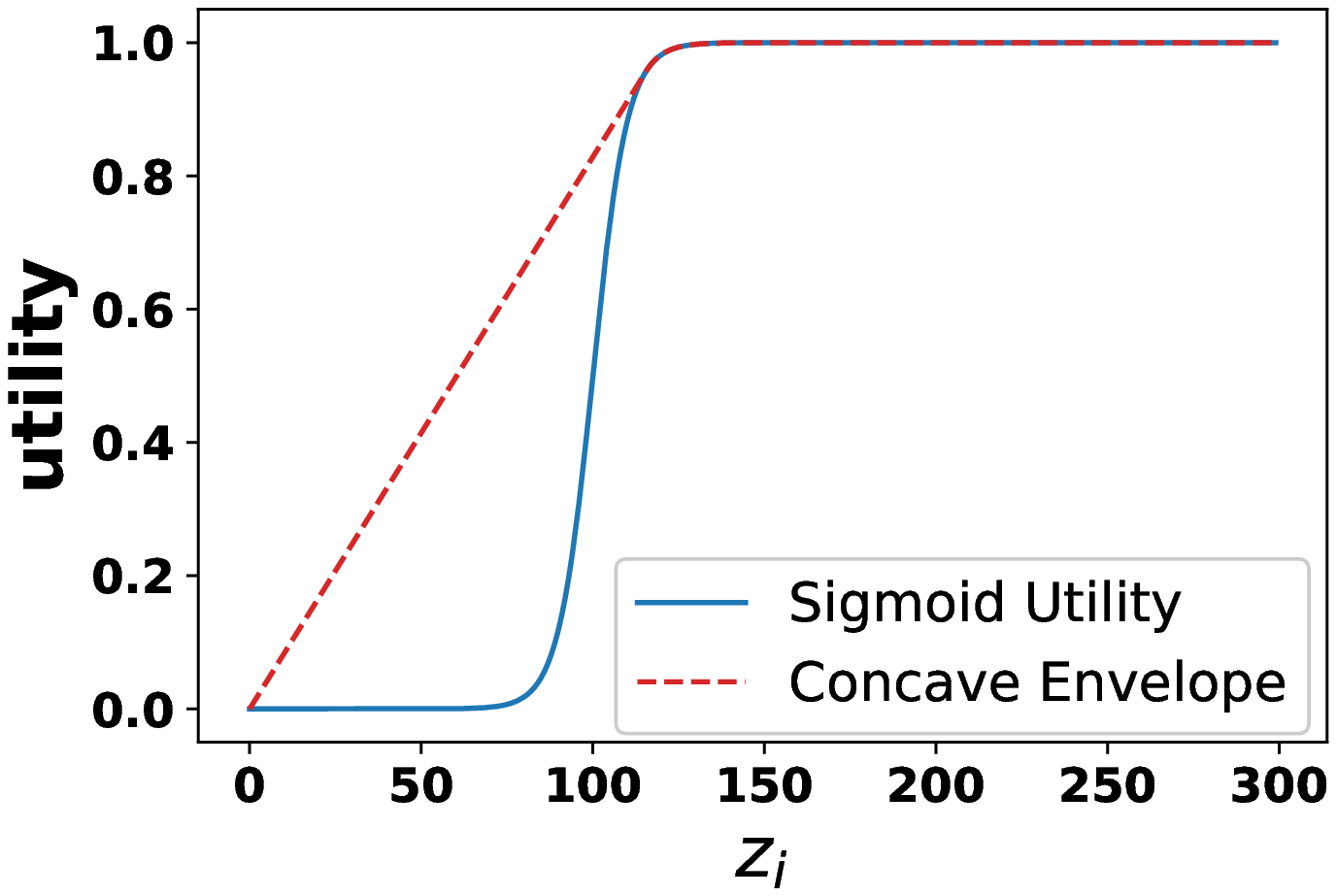}
\vspace{-20pt}
\caption{} 
\label{fig:0b}
\end{subfigure}
\begin{subfigure}[b]{0.32\linewidth}
\includegraphics[height=4.5cm,width=\linewidth]{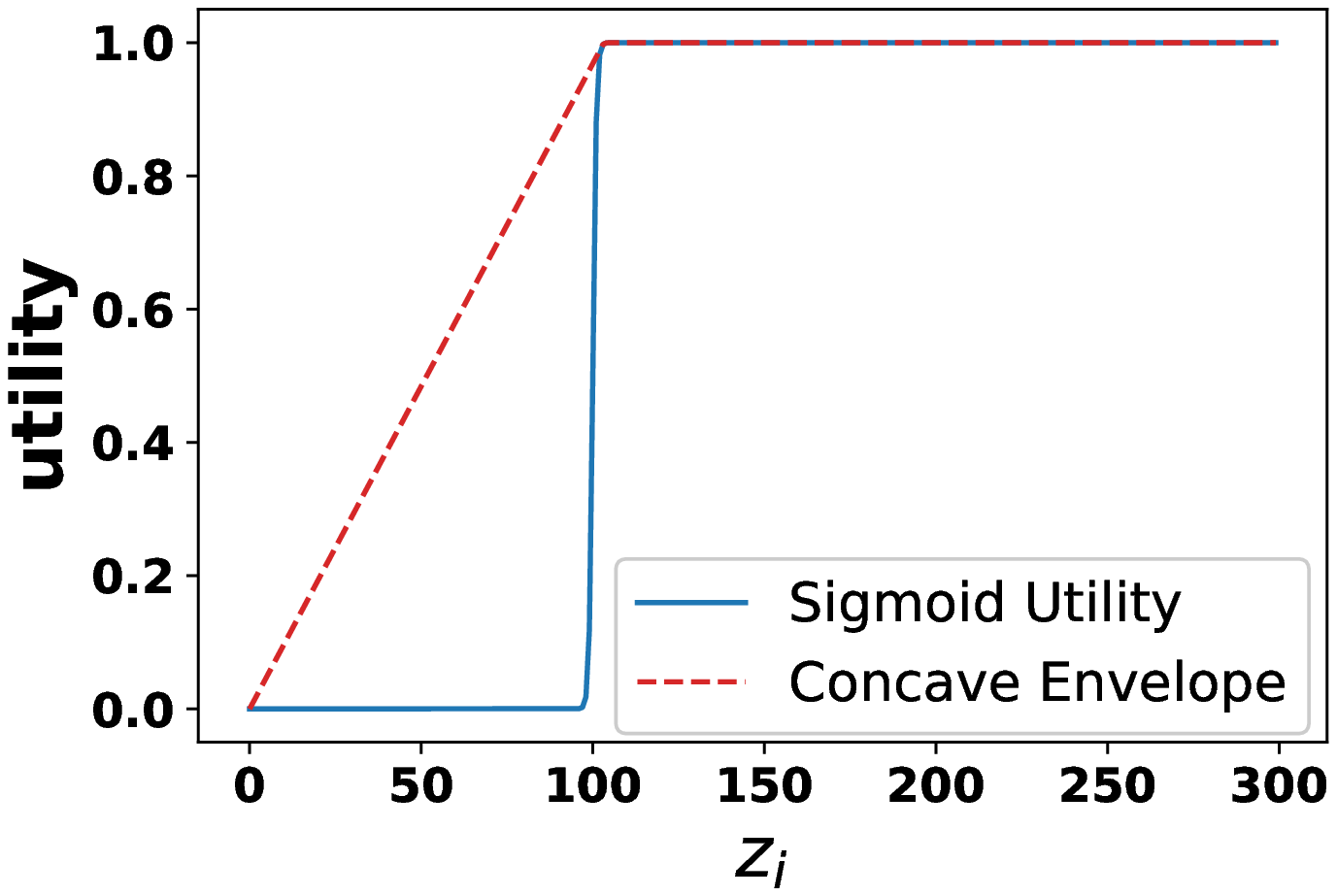}
\vspace{-20pt}
\caption{} 
\label{fig:0c}
\end{subfigure}
\vspace{-5pt}
\caption{\small{Concave Envelopes of sigmoid utility functions with $k_{\mathbb{c}(\cdot)} = 100$ and (a) $t^z_{\mathbb{c}(\cdot)} = 0.02$, (b) $t^z_{\mathbb{c}(\cdot)} = 0.2$ and (c) $t^z_{\mathbb{c}(\cdot)} = 2$.}} 
\label{fig:0}
\end{figure}

To concavify $\bar{\bm{P}}$,
we replace every summand of $u_m$ with its tightest concave envelope, i.e., the pointwise infimum over all concave functions that are greater or equal. For the sigmoid function 
$u_{i}(z_i)$ the concave envelope, $\hat{u}_{i}(z_i)$, has a  closed form given by
\[ 
    \hat{u}_{i}(z_i) =   
    \left\{
\begin{array}{ll} 
      u_{i}(0)+\frac{u_{i}(w) - u_{i}(0)}{w}z_i & \scriptstyle 0 \le z_i \le w\\
      u_{i}(z_i) & \scriptstyle w \le z_i   \\
\end{array} 
\right.,
\]
for some $w > k_i$ which can be found easily by bisection \cite{udell2013maximizing}. Fig. \ref{fig:0} depicts the concavification of the aforementioned sigmoid functions for $k_{\mathbb{c}(\cdot)} = 100$ and three different values for $t^z_{\mathbb{c}(\cdot)}$. Note that for the lowest $t^z_{\mathbb{c}(\cdot)}$ (elastic traffic) we get the best approximation whilst for the largest (inelastic traffic/tight QoS prerequisites) we get the worst.

\par

To exploit the closed form of the envelope $\hat{u}_{i}(z_i)$, instead of problem $\bar{\bm{P}}$, we will concavify the equivalent problem:  
\begin{equation*}
\begin{aligned} 
\textbf{($\bm{\tilde{P})}$: $\quad$} \max_{\bm{r}_m, \bm{x_m}, \bm{z}_m,} \quad &   \sum_{i \in \mathcal{U}_m} f_{i}(\bm{r}_i, z_i) - \bm{c}^T \bm{x}_{m} - \lambda_m \|\bm{x}_m\|_2^2, \\
\textrm{s.t.} \quad & (\bm{r}_i, z_i) \in S_{i},  \quad \forall i \in \mathcal{U}_m\\
  \quad & \bm{x}_m \succeq \sum_{i \in \mathcal{U}_m} \bm{r}_i \\
\end{aligned}    
\end{equation*}
where $S_i := \{ (\bm{r}_i, z_i) : \bm{r}_i\succeq \bm{0}, z_i = \bm{\beta}_i^T\bm{r}_i  \ \} $ and 
$f_{i}(\bm{r}_i, z_i) := u_{i}(z_i)$ with domain $S_{i}$. 
The following lemma uses the concave envelope of the sigmoid function $u_{i}(z_i)$, to compute the concave envelope of $f_{i}(\bm{r}_i, z_i)$ and hence the concavification of the problem \textbf{$\bm{\tilde{P}}$}. Its proof is based on the definition of the concave envelope and is omitted for brevity.

\begin{lemma}
The concave envelope of the function $ f_{i}(\bm{r}_i, z_i) :=   \frac{e^{t^{z}_{\mathbb{c}(i)} (z_i-k_{\mathbb{c}(i)})}}{1+e^{t^{z}_{\mathbb{c}(i)} (z_i-k_{\mathbb{c}(i)})}}$ with domain $S_{i}$, $\hat{f}_{i}(\bm{r}_i, z_i)$, has the following closed form (with domain $S_{i}$):   
\[ 
    \hat{f}_{i}(\bm{r}_i, z_i) =   \hat{u}_{i}(z_i), \quad \forall (\bm{r}_i, z_i) \in S_{i}.
\]
\end{lemma}

Therefore, SP $m$ can concavify $\tilde{\bm{P}}$ as follows:
\begin{equation*}
\begin{aligned} 
\textbf{($\bm{\hat{P})}$: $\quad$} \max_{\bm{r}_m, \bm{x_m}, \bm{z}_m} \quad &   \sum_{i \in \mathcal{U}_m} \hat{f}_{i}(\bm{r}_i, z_i) - \bm{c}^T \bm{x}_{m} - \lambda_m \|\bm{x}_m\|_2^2\\
\textrm{s.t.} \quad & (\bm{r}_i, z_i) \in S_{i},  \quad \forall i \in \mathcal{U}_m\\
  \quad & \bm{x}_m \succeq \sum_{i \in \mathcal{U}_m} \bm{r}_i \\
\end{aligned}    
\end{equation*}
Note that \textbf{$\bm{\hat{P}}$} is strongly concave and thus admits a unique maximizer. 
Let the solution and the optimal point of problem \textbf{$\bm{\hat{P}}$} be $\hat{\psi}_m^{*}$ and $(\hat{\bm{x}}^{*}_m, \hat{\bm{r}}^{*}_m)$ respectively. Ultimately, we would like to compare the solution of the concavified $\bm{\hat{P}}$ with the one of the original problem $\bm{{P}}$. Towards that direction, we first define the \textit{nonconcavity of a function} as follows \cite{udell2016bounding}:

\begin{definition}[Nonconcavity of a function] We define the nonconcavity $\rho(f)$ of a function $f:S \rightarrow \mathbb{R}$ with domain $S$, to be 
\begin{equation*}
    \rho(f) = \sup_x (\hat{f}(x) - f(x)).
\end{equation*}
\end{definition}
Let $\mathcal{F}$ denote a set of possibly non-concave functions. Then define $\rho_{ [j] }(\mathcal{F})$ to be the $j$th largest of the nonconcavities of the functions in $\mathcal{F}$. The theorem below, summarizes the main result of this section, which is that every SP can solve the concavified $\bm{\hat{P}}$ instead of the original $\bm{P}$, since the former provides a constant bound approximation of the latter.
Recall that $\Psi_m(\bm{\hat{r}}_m^*, \bm{\hat{x}}^*_{m})$ is the profit of SP $m$, evaluated at the solution of $\bm{\hat{P}}$ and that $K$ is the number of the NPs.



\begin{theorem}
\label{mainth}
Let $(\bm{r}_m^*, \bm{x}_m^*)$ and $(\bm{\bar{r}}_m^*, \bm{\bar{x}}_m^*)$ be solutions of Problems $\bm{P}$ and $\bm{\bar{P}}$ respectively. Moreover, let $\hat{\mathcal{F}} := \{u_{i}\}_{i \in \mathcal{U}_m}$. Then,
$$\psi_m^* - \epsilon - \delta_1(\lambda_m) \le  \Psi_m(\bm{\hat{r}}_m^*, \bm{\hat{x}}^*_{m}) \le \psi_m^* + \delta_2(\lambda_m),$$
where $\delta_1(\lambda_m) := \lambda_m(\|\bm{x}_m^*\|_2^2 - \|\bm{\hat{x}}_m^*\|_2^2)$, $\delta_2(\lambda_m) := \lambda_m(\|\bm{\hat{x}}_m^*\|_2^2 - \|\bm{\bar{x}}_m^*\|_2^2)$ and $\epsilon = \sum_{j = 1}^{K} \rho_{[j]}(\hat{\mathcal{F}})$.
\begin{IEEEproof}

Note that $\bar{\psi}_m^{*}$ is also given by solving $\bm{\tilde{P}}$ and that $(\bm{\hat{r}}^*_m, \bm{\hat{x}}^*_m)$ with the corresponding optimal value $\hat{\psi}_m^{*}$, are given by solving $\bm{\hat{P}}$. Therefore, from \cite[Th. 1]{udell2016bounding}, we have that $$\bar{\psi}_m^* - \sum_{j = 1}^{K} \rho_{[j]}(\hat{\mathcal{F}}) \le u_m(\bm{\hat{r}}^*_m) - \bm{c}^T \bm{\hat{x}}^*_m - \lambda_m \| \bm{\hat{x}}^*_m\|_2^2 \le \bar{\psi}_m^* $$
The result follows from Lemma \ref{lemma_regularization_optimality}.
\end{IEEEproof}
\end{theorem}

\begin{remark}
\label{remark1}
The values of $\delta_1$ and $\delta_2$ decrease as $\lambda_m$ decreases and hence for small regularization penalties they can get arbitrarily close to zero.
\end{remark}

\begin{remark}
\label{remark2}
The approximation error, $\epsilon$, depends on the $K$ greatest nonconcavities of the set $\{u_{i}\}_{i \in \mathcal{U}_m}$. There are two conditions that ensure negligible approximation error, i.e.,  $\epsilon << \psi^*_m$: i) the end-users have concave utility functions (in that case $\epsilon \rightarrow 0$) or, ii) the market is profitable enough for every SP $m$ and hence $\psi^*_m >> K$. Condition ii) makes the error negligible since $\epsilon \le K$, and it can be satisfied for example when the supply of the market, $\bm{C}$, is sufficiently large. 
\end{remark}

Theorem \ref{mainth}, implies that every SP can solve Problem $\bm{\hat{P}}$, which is a concave program with a unique solution, to find an approximate solution to $\bm{P}$. This observation fosters the convergence analysis of the proposed auction in Section \ref{sec:auction}.

\section{Network Slicing Market Cycle}
In this section, we study the evolution of the network slicing market using an iterative model that consists of $5$-step cycles. 
We refer to the following sequence of steps as a market cycle:
\begin{enumerate}[S1.]
    \item $|\mathcal{U}_m|$ prospective users appear to every SP $m$. 
    \item The vector $\bm{x_m}$, i.e., the distribution of the resources from the NPs to SP $m$ is determined for every $m$. To achieve that in a distributed fashion, an auction between the SPs and the NPs should be realized.
    \item Given $\bm{x_m}$, each SP $m$ determines the vectors $\bm{r}_i$ and hence the amount of resources $z_i$ for every user $i \in \mathcal{U}_m$ (intra-slice resource allocation). 
    \item After receiving the resources, each user $i$ determines and reports to the SP whether the QoS received was enough or not to complete its application. 
    \item The SPs exploit the responses of their users, to estimate their private parameters and hence to distribute the resources more efficiently in the next cycle.

\end{enumerate}
It is important for the vector $\bm{x}_m$ to be determined before the intra-slice resource allocation, since the first serves as the capacity in the resources available to SP $m$. 
In the following, we expand upon each (non-trivial) step of the market cycle.


\subsection{Step S2 - Clock Auction for the Network Slicing Market}
\label{sec:auction}



In this section, we develop and analyze a clock auction between the SPs and the NPs, that converges to a market's equilibrium. Specifically, we describe the goal (Section \ref{sec:au_goal}), the steps (Section \ref{sec:au_steps}), and the convergence (Section \ref{sec:au_convergence}) of the auction.

\subsubsection{Auction Goal}
\label{sec:au_goal}


Note that the solutions of the problems $\bm{{P}}$ and $\bm{\hat{P}}$ appear to be a function of the prices $c_1, \dots, c_{K}$. Let the demand of SP $m$, given the price vector $\bm{c}$, be denoted as $\bm{{x}}^{*}_m(\bm{c})$ or $\bm{\hat{x}}^{*}_m(\bm{c})$ depending on whether SP $m$ uses Problem $\bm{{P}}$ or $\bm{\hat{P}}$ to ask for resources. Let also $\bm{{r}}^{*}_m(\bm{c})$ and $\bm{\hat{r}}^{*}_m(\bm{c})$ be optimal intra-slice resource allocation vectors respectively. Hence,  $(\bm{{r}}^{*}_m(\bm{c}), \bm{{x}}^{*}_m(\bm{c}))$ and $(\bm{{\hat{r}}}^{*}_m(\bm{c}), \bm{{\hat{x}}}^{*}_m(\bm{c}))$ are maximizers of $\bm{P}$ and $\bm{\hat{P}}$ respectively (given $\bm{c}$). Since Problem $\bm{P}$ may admit multiple solutions, let the set $\mathcal{D}_m(\bm{c})$ be defined as  $$\mathcal{D}_m(\bm{c}) := \bigg\{ \bm{x}_m^{*} : \{ \exists \bm{r}_m^{*} : \{ \Psi_m(\bm{r}_m^{*}, \bm{x}_m^{*}) = \psi^*_m  \text{ given } \bm{c} \} \bigg\}.$$




We define a Competitive equilibrium as follows:

\begin{definition}[Competitive equilibrium]
Competitive equilibrium of the Network Slicing Market is defined to be any price vector $\bm{c}^{\dagger}$ and allocation of the resources of the NPs $\bm{{x}}^{\dagger}$, such that:
\begin{enumerate}[i.]
    
    \item $\bm{{x}}_m^{\dagger} \in \mathcal{D}_m(\bm{c}^\dagger)$ for every SP $m$, 
    and
    \item $ \bm{C} = \sum_{m \in \mathcal{M}} \bm{{x}}_m^{\dagger}$ (the demand equals the supply).
\end{enumerate}
    
\end{definition}


Note that in a competitive equilibrium, every SP $m$ gets resources that could maximize its profit given the price vector. Because a competitive equilibrium sets a balance between the interests of all participants, it appears to be the settling point of the markets in economic analysis \cite{bichler2021walrasian, shen2018first}. Nevertheless, since the SPs' demands are expressed by solving a non-concave program, we define an $\epsilon$-competitive equilibrium which will be the ultimate goal of the proposed clock auction.


\begin{definition}[$\epsilon$-Competitive equilibrium]
$\epsilon$-Competitive equilibrium of the Network Slicing Market is defined to be any price vector $\bm{\hat{c}}^{\dagger}$ and allocation of the resources of the NPs $\bm{{\hat{x}}}^{\dagger}$, such that:
\begin{enumerate}[i.]
    \item For every SP $m$, there exists an $\epsilon \ge 0$ and a feasible intra-slice resource allocation vector $\bm{\hat{r}}_m^\dagger$ (given $\bm{\hat{x}}^\dagger_{m}$), such that: $$\psi_m^* - \epsilon  \le  \Psi_m(\bm{\hat{r}}_m^\dagger, \bm{\hat{x}}^\dagger_{m}) \le \psi_m^* + \epsilon, \text{ and}$$
    \item $ \bm{C} = \sum_{m \in \mathcal{M}} \bm{{\hat{x}}}_m^{\dagger}$ (the demand equals the supply).
\end{enumerate}
\end{definition}

Observe that the first condition of the above definition ensures that every SP is satisfied (up to a constant) with the obtained resources in a sense that it operates close to its maximum possible profit. From Theorem \ref{mainth}, note that if there exists a price vector $\bm{\hat{c}}^\dagger$ such that $ \bm{C} = \sum_{m \in \mathcal{M}} \bm{{\hat{x}}}_m^{*}(\bm{\hat{c}^\dagger})$, then the prices in $\bm{\hat{c}}^\dagger$ with the allocation $\bm{{\hat{x}}}^{\dagger} := \bm{{\hat{x}}}^{*}(\bm{\hat{c}^\dagger})$ form an $\epsilon$-competitive equilibrium. Finding such a price vector, is the motivation of the proposed clock auction. For the rest of the paper we make the following assumption:
\begin{assumption}
\label{as:demandofsps}
The SPs calculate their demand and intra-resource allocation by solving Problem $\bm{\hat{P}}$.
\end{assumption}
This is a reasonable assumption since in Theorem \ref{mainth} and the corresponding Remarks \ref{remark1} and \ref{remark2}, we proved that by solving a (strictly) concave problem, every SP can operate near its optimal profit. Therefore, for the rest of the paper, we call $\hat{x}^*_m(\bm{c})$, the demand of SP $m$ given the prices $\bm{c}$.


\subsubsection{Auction Description}
\label{sec:au_steps}

We propose the following clock auction that converges to an $\epsilon$-competitive equilibrium of the Network Slicing market (Theorem \ref{th:convergence}). As we will prove in Theorem \ref{th:uniqueness}, this equilibrium is robust since the convergent price vector is the unique one that clears the market, i.e., makes the demand to equal the supply. 

\begin{enumerate}[i.]
    \item An auctioneer announces a price vector $\bm{c}$, each component of which corresponds to the price that an NP sells a unit of its resources.
    \item The bidders (SPs) report their demands.
    \item If the aggregated demand received by an NP is greater than its available supply, the price of that NP is increased and vice versa. In other words, the auctioneer adjusts the price vector according to Walrasian tatonnement.
    \item The process repeats until the price vector converges.
\end{enumerate}

Note that the components of the price vector change simultaneously and independently. Hence different brokers can cooperate to jointly clear the market efficiently in a decentralized fashion \cite{iosifidis2014double}. Let the excess demand, $\mathcal{Z}(\bm{c})$,  be the difference between the aggregate demand and supply: $\mathcal{Z}(\bm{c}) = -\bm{C} + \sum_{m \in \mathcal{M}} \bm{\hat{x}}^{*}_m(\bm{c})$. In Walrasian tatonnement, the price vector adjusts in continuous time according to excess demand as $\dot{\bm{c}} =  f(\mathcal{Z}(\bm{c}(t)))$, where $f$ is a continuous, sign-preserving transformation \cite{ausubel2004auctioning}. For the rest of the paper, we set $f$ to be the identity function and thus $\dot{\bm{c}} =  \mathcal{Z}(\bm{c}(t))$. In auctions based on Walrasian tatonnement, the payments are only valid after the convergence of the mechanism \cite{courcoubetis2003pricing}. 

\subsubsection{Auction Convergence}
\label{sec:au_convergence}

Towards proving the convergence of the auction, we provide the lemma below which proves that the concavified version of the intra-slice resource allocation problem $\bm{IN-SL}$, can be thought of as a concave function. The proof is ommitted as a direct extension of \cite{qin2020network} and \cite{boyd2004convex}.

\begin{lemma}
The function $U_m(\bm{x}_m)$ shown below is concave.
\begin{equation}
\label{demand_utility}
\begin{aligned} U_m(\bm{x}_m) :=
 \max_{\bm{r}_m, \bm{z}_m} \quad &   \sum_{i \in \mathcal{U}_m} \hat{f}_{i}(\bm{r}_i, z_i) \\
\textrm{s.t.} \quad & (\bm{r}_i, z_i) \in S_{i},  \quad \forall i \in \mathcal{U}_m\\\
  \quad & \bm{x}_m \succeq \sum_{i \in \mathcal{U}_m} \bm{r}_i \\
\end{aligned}
\end{equation}
\end{lemma}
Using the function $U_m$, we can rewrite Problem $\bm{\hat{P}}$ as 
$$
\begin{aligned} 
\max_{\bm{x_m} \succeq 0} \quad &  U_m(\bm{x}_m) - \lambda_m - \bm{c}^T \bm{x}_{m}  \|\bm{x}_m\|_2^2\\
\end{aligned}.
$$
The following theorem studies the convergence of the auction.
\begin{theorem}
\label{th:convergence}
Starting from any price vector $\bm{c}_{init}$, the proposed clock auction converges to an $\epsilon$-competitive equilibrium.

\begin{IEEEproof}
The proof relies on a global stability argument similarly to \cite{ausubel2004auctioning, bichler2021walrasian}. Let $\mathcal{V}_m( \cdot )$ denote $m$'s net indirect utility function:
$$\mathcal{V}_m(\bm{c}) = \max_{\bm{x_m} \succeq 0 } \quad \{ U_m(\bm{x}_m) - \lambda_m \lVert \bm{x}_m \rVert_2^2 - \bm{c}^T \bm{x}_{m} \}.$$ Let a candidate Lyapunov function be $\mathcal{V}(\bm{c}) := \bm{c}^T \bm{C} + \sum_{m \in \mathcal{M}} \mathcal{V}_m(\bm{c})$. To study the convergence of the auction we should find the time derivative of the above Lyapunov function:
\[
\label{lyapunov_dot}
\scriptstyle
\dot{ \mathcal{V}}(\bm{c}) = \dot{c} \cdot \Big(  \bm{C}^T + \sum_{m \in \mathcal{M}  }  \frac{d}{d\bm{c}}\Big\{ \max_{\bm{x_m} \succeq 0} \{  U_m(\bm{x}_m) - \lambda_m \lVert \bm{x}_m \rVert_2^2 - \bm{c}^T \bm{x}_{m} \} \Big\}  \Big).
\]
Hence, we deduce that:
\begin{equation*}
\dot{ \mathcal{V}}(\bm{c}) =  \Big(  \bm{C}^T + \sum_{m \in \mathcal{M}  } \{  - \bm{\hat{x}}_m^{*T}(\bm{c}) \}  \Big) \cdot \dot{c} = - \mathcal{Z}^{T}(\bm{c}(t)) \cdot  \mathcal{Z}(\bm{c}(t)).
\end{equation*}
The above holds true since the function $h(\bm{x}_m) := U_m(\bm{x}_m) - \lambda_m \lVert \bm{x}_m \rVert_2^2$, has as concave conjugate the function (see \cite{boyd2004convex})
$$h^{*} (\bm{s}) = \max_{\bm{x_m} \succeq 0} \{ 
h(\bm{x_m}) - \bm{c}^T \bm{x}_{m} \},$$ and hence $\nabla h^{*}(\bm{s}) = 
arg\max_{\bm{x_m} \succeq 0} \{ U_m(\bm{x}_m) - 
\lambda_m \lVert \bm{x}_m \rVert_2^2 - \bm{c}^T 
\bm{x}_{m} \}.$
Therefore, $\mathcal{V}(\cdot)$ is a decreasing function of time and converges to its minimum. Note that in the convergent point the supply equals the demand for every NP.
\end{IEEEproof}
\end{theorem}
The market might admit multiple $\epsilon$-competitive equilibria. Nevertheless, the equilibrium point that the clock auction converges is robust in the following sense: given 
Assumption \ref{as:demandofsps}, the price vector that clears the market is unique. Therefore, essentially, in Theorem \ref{th:convergence} we proved that the proposed clock auction converges to that unique price vector. This is formally proposed by the following theorem.

\begin{theorem}
\label{th:uniqueness}
There exists a unique price vector $\bm{c}^{\dagger}$ such that $ \sum_{m \in \mathcal{M}}  \bm{\hat{x}}_m^{*}(\bm{c^\dagger})  = \bm{C}$.
\end{theorem}

Towards proving Theorem \ref{th:uniqueness} we provide Lemmata \ref{uniqueness_x} and \ref{warp_lemma}. First, we show that if a component in the price vector changes, the demand of an SP who used to obtain resources from the corresponding NP, should change as well.

\begin{lemma}
\label{uniqueness_x}
For two distinct price vectors $\bm{c}$, $\bar{\bm{c}}$ with $\exists k : c_k \neq \bar{c}_k$, it holds true that \[
 \bm{\hat{x}}_m^{*}(\bm{c}) = \bm{\hat{x}}_m^{*}(\bm{\bar{c}}) \Rightarrow \hat{x}_{(m,k)}^{*}(\bm{c}) = \hat{x}_{(m,k)}^{*}(\bar{\bm{c}}) = 0. 
\]
\begin{IEEEproof}
Let such price vectors, $\bm{\bar{c}}$ and $\bm{c}$, with $c_k \neq \bar{c}_k$. Since $\bm{\hat{x}}_m^{*}(\bm{c})$  is the optimal point of problem $\bm{\hat{P}}$ given $\bm{{c}}$, applying KKT will give:
\small
\begin{equation}
\label{eqcam}
    \hat{x}_{(m,k)}^{*}(\bm{c}) = 0 \quad or \quad \frac{\partial\{ U_m(\bm{x}_m) - \lambda_m \lVert \bm{x}_m  \rVert_2^2\}}{\partial  x_{(m,k)}}\bigg|_{\bm{\hat{x}}_m^{*}(\bm{c})} = c_k.
\end{equation}
  \normalsize
However, $\bm{\hat{x}}_m^{*}(\bm{\bar{c}})$ is optimal for $\bm{\hat{P}}$ given $\bm{\bar{c}}$. Employing a similar equation as \eqref{eqcam} proves that if $\bm{\hat{x}}_m^{*}(\bm{c}) = \bm{\hat{x}}_m^{*}(\bm{\bar{c}})$ then it can only hold that $\hat{x}_{(m,k)}^{*}(\bm{c}) = \hat{x}_{(m,k)}^{*}(\bar{\bm{c}}) = 0$.




\end{IEEEproof}
\end{lemma}

\begin{definition}[WARP property]
The aggregate demand function satisfies the \textit{Weak Axiom of Revealed Preferences (WARP)}, if for different price vectors $\bm{c}$ and $\bar{\bm{c}}$, it holds that:
\small
\[
\bm{c}^T \cdot \sum_{m \in \mathcal{M}} \bm{\hat{x}}_m^{*}(\bar{\bm{c}}) \le \bm{c}^T \cdot \sum_{m \in \mathcal{M}} \bm{\hat{x}}_m^{*}(\bm{c}) \Rightarrow  \qquad \qquad \qquad \qquad
\]
\[ \qquad \qquad \qquad \qquad
\bar{\bm{c}}^T \cdot \sum_{m \in \mathcal{M}} \bm{\hat{x}}_m^{*}(\bar{\bm{c}}) <
\bar{\bm{c}}^T \cdot \sum_{m \in \mathcal{M}} \bm{\hat{x}}_m^{*}(\bm{c}) 
\]
\normalsize
\end{definition}

\begin{lemma}
\label{warp_lemma}
The aggregate demand function satisfies the WARP for distinct price vectors $\bm{c}$, $\bm{\bar{c}}$ such that $\sum_{m \in \mathcal{M}} \bm{\hat{x}}_m^{*}(\bm{c}) \succ \bm{0} $ and $\sum_{m \in \mathcal{M}} \bm{\hat{x}}_m^{*}(\bm{\bar{c}}) \succ \bm{0} $.
\begin{IEEEproof}
Since $\bm{c} \neq \bm{\bar{c}}$ then $\exists k \in \mathcal{K}: c_k \neq \bar{c}_k$. Furthermore, we have that $\sum_{m \in \mathcal{M}} \bm{\hat{x}}_m^{*}(\bm{c}) \succ \bm{0} $ and hence $\exists m_1 \in \mathcal{M}$ such that $\hat{x}_{m_1, k}^{*}(\bm{c}) > 0$. Using Lemma \ref{uniqueness_x} we conclude that $\bm{\hat{x}}_{m_1}^{*}(\bm{c}) \neq \bm{\hat{x}}_{m_1}^{*}(\bm{\bar{c}})$. Hence, since Problem $\bm{\hat{P}}$ admits a unique global maximum we have that:
\small
\[
 \sum_{m \in \mathcal{M}} \Big\{ U_m(\bm{\hat{x}}_m^{*}(\bm{c})) - \lambda_m \lVert \bm{\hat{x}}_m^{*}(\bm{c}) \rVert_2^2  - \bm{c}^T \cdot \bm{\hat{x}}_m^{*}(\bm{c}) \Big\} > \qquad \qquad 
 \]
 \normalsize
  \small 
 \[
 \quad \quad
 \sum_{m \in \mathcal{M}} \Big\{ U_m(\bm{\hat{x}}_m^{*}(\bm{\bar{c}})) - \lambda_m \lVert \bm{\hat{x}}_m^{*}(\bm{\bar{c}}) \rVert_2^2  - \bm{c}^T \cdot \bm{\hat{x}}_m^{*}(\bar{\bm{c}}) \Big\} 
\]
 \normalsize

Now, the above combined with the WARP hypothesis,
\[
 \sum_{m \in \mathcal{M}} \bm{c}^T \cdot \bm{\hat{x}}_m^{*}(\bar{\bm{c}}) \le  \sum_{m \in \mathcal{M}} \bm{c}^T \cdot \bm{\hat{x}}_m^{*}(\bm{c}), 
\]
gives:
\small
\[
\sum_{m \in \mathcal{M}} \Big\{ U_m(\bm{\hat{x}}_m^{*}(\bm{c})) - \lambda_m \lVert \bm{\hat{x}}_m^{*}(\bm{c}) \rVert_2^2 \Big\}  >  \qquad \qquad \qquad \qquad
\]
\normalsize
\small
\begin{equation}
\label{temp2}
\small
\qquad \qquad
\sum_{m \in \mathcal{M}} \Big\{  U_m(\bm{\hat{x}}_m^{*}(\bm{\bar{c}})) - \lambda_m \lVert \bm{\hat{x}}_m^{*}(\bm{\bar{c}}) \rVert_2^2 \Big\}.
\end{equation}
\normalsize

The result follows by switching the roles of $\bm{c}$ and $\bm{\bar{c}}$ and combine the inequalities.



\end{IEEEproof}
\end{lemma}

We can now prove Theorem \ref{th:uniqueness} as follows.


\begin{IEEEproof}[proof of Theorem \ref{th:uniqueness}]
Towards a contradiction, assume that there exist two distinct (non-zero) price vectors $\bm{c}$ and $\bar{\bm{c}}$ that satisfy
$\sum_{m \in \mathcal{M}}  \bm{\hat{x}}_m^{*}(\bar{\bm{c}}) = \sum_{m \in \mathcal{M}}  \bm{\hat{x}}_m^{*}(\bm{c})  = \bm{C}$ and thus
\begin{equation}
    \label{warp2}
    \bm{c}^T  \cdot \Big( \sum_{m \in \mathcal{M}}  \bm{\hat{x}}_m^{*}(\bar{\bm{c}}) - \sum_{m \in \mathcal{M}}  \bm{\hat{x}}_m^{*}(\bm{c}) \Big) = 0. 
\end{equation}
Therefore, from Lemma \ref{warp_lemma} we know that:
\begin{equation}
    \label{warp4}
    \bar{\bm{c}}^T \cdot \sum_{m \in \mathcal{M}} \bm{\hat{x}}_m^{*}(\bar{\bm{c}}) < \bar{\bm{c}}^T \cdot \sum_{m \in \mathcal{M}} \bm{\hat{x}}_m^{*}(\bm{c}), 
\end{equation}
which is a contradiction because of the hypothesis. 

\end{IEEEproof}

\par

\begin{remark}
\label{remark_concave}
Theorems \ref{th:convergence} and \ref{th:uniqueness} together with Remarks \ref{remark1} and \ref{remark2} imply that if the users' traffic is elastic, or the total capacity $\bm{C}$ of the NPs is sufficiently large, the clock auction converges monotonically to the unique competitive equilibrium of the market. 
\end{remark}

At the end of step $S2$, the final price vector $\bm{\hat{c}}^{\dagger}$ and the final demands of each SP $m$, $\bm{\hat{x}}_m^{*}$, have been determined. 

\subsection{Intra-Slice Resource Allocation \& Feedback (Steps S3, S4)}
At the beginning of step $S3$, every SP $m$ is aware of the convergent point $\bm{\hat{x}}_m^{*}$ and hence it can allocate the resources either by solving the sigmoid program $\bm{IN-SL}$, or by using the convergent approximate solution, $\bm{\hat{r}}_m^*$. 
At that step, an SP can also determine whether it will \textit{overbook} network resources. Overbooking, is a common practice in airlines and hotel industries and is now being used in the network slicing problem \cite{salvat2018overbooking}, \cite{marquez2019resource}. This management model allocates the same resources to users of the network expecting that not everyone uses their booked capacity. In that case, SP $m$ solves Problem $\bm{IN-SL}$ whilst setting increased obtained resources, $\bm{x}_m^{ov} = \bm{\hat{x}}_m^{*} + \bm{\alpha} \% \circ \bm{\hat{x}}_m^{*}$, for a relatively small positive $\bm{\alpha}$. Here, $\circ$ denotes the component-wise multiplication operator.
\par
During the step S4 of the cycle, each user $i$, receives their resources $\bm{r}_i$, and provide feedback on whether it was satisfied or not. In the next step, the SPs can use the these responses to learn the private parameters of the different service classes.

\subsection{Learning the Parameters (Step S5)}
\label{sec:learning}
At the final step of the cycle, the SPs exploit the data they obtained to learn the private parameters of their users. In that fashion, the market ''learns" its equilibrium. For the rest of the paper, for generality, we assume the pricing mechanism introduced in Section \ref{sec:pricingmechanism}. Therefore, for every user $i$, the SPs get to know whether it is satisfied by the pair of resources-price $(z_i, p_i)$. A Bayesian inference model needs the data, a model for the private parameters and a prior distribution.
\par
\textbf{Model:} The observed data is the outcome of the Bernoulli variables $sat_i|\theta_{\mathbb{c}(i)} \sim Bernoulli(P[sat_i])$ for every user $i$, where $\theta_{\mathbb{c}(i)} = (t_{\mathbb{c}(i)}^{p}, b_{\mathbb{c}(i)}, t_{\mathbb{c}(i)}^{z}, k_{\mathbb{c}(i)})$ is the tuple of the private parameters that we want to infer. 
\textbf{Prior:} Let the prior distribution for every parameter of $\theta_{\mathbb{c}(i)}$
have probability density functions $\pi_{t_{\mathbb{c}(i)}^{p}}(\cdot), \pi_{b_{\mathbb{c}(i)}}(\cdot), \pi_{t_{\mathbb{c}(i)}^{z}}(\cdot)$ and $\pi_{k_{\mathbb{c}(i)}}(\cdot)$ respectively. The SPs infer the private parameters $\theta_{\mathbb{c}(i)}$ for each service class using the Bayes rule separately:
$
    p(\theta_{\mathbb{c}(i)} | data) \propto L_n(data | \theta_{\mathbb{c}(i)}) \pi(\theta_{\mathbb{c}(i)}),
$
where $p(\theta_{\mathbb{c}(i)} | data)$ is the posterior distribution of $\theta_{\mathbb{c}(i)}$, $ L_n(data | \theta_{\mathbb{c}(i)})$ is the likelihood of the data given our model and $\pi(\theta_{\mathbb{c}(i)})$ is the prior distribution.
Assuming independent private parameters, $\pi(\theta_{\mathbb{c}(i)})$ is the product of the distinct prior distributions, and for each class $c$ we have that:
\begin{equation*}
L_n(data | \theta_{\mathbb{c}(i)}) = \prod_{i \in C_c^m} P[sat_i]^{f_i} (1 - P[sat_i])^{1-f_i},
\end{equation*}
where $f_i$ is $1$ when user $i$ is satisfied and $0$ when not. 

\par
The SPs can use \textit{Marcov Chain Monte Carlo (MCMC)} with \textit{Metropolis Sampling}, to find the posterior distribution after each market cycle. As the market evolves, the SPs exploit the previous posterior distributions to find better priors for the next cycle.


\section{Centralized Solution}
In case there exists a centralized entity that knows the utility function of every SP, it can optimize the social welfare, i.e., the summation of the utility functions of the service and the network providers. 
This centralized problem can be formulated as follows:

\begin{equation*}
      \begin{aligned} 
\textbf{(SWM): $\quad$}  \max_{\bm{r_m}} \quad &  \sum_{m \in \mathcal{M}} u_m(\bm{r_m})\\
\textrm{s.t.} \quad & \bm{r}_i \succeq \bm{0}, \quad \forall i \in \mathcal{U}_m \\
  & \sum_{m \in \mathcal{M}} \sum_{i \in U_m} \bm{r_{i}} \preceq \bm{C} \\
\end{aligned}    
\end{equation*}
The $\bm{SWM}$ problem, can be solved with any chosen positive approximation error, using the framework of sigmoidal programming \cite{udell2013maximizing}.

\section{Numerical Results}
\subsection{Auction Convergence \& Parameter Tuning}

In this section we study the convergence of the clock auction, as well as the impact that  the various parameters have on its behavior. For this simulation, we assume a small market with $3$ NPs with capacities $C_1 = 850, C_2 = 750, C_3 = 755$ and $5$ SPs with $6$ users and $3$ distinct service classes each. The users' private parameters are set as follows: for an $i$ in the first class $t_{\mathbb{c}(i)}^z = t_{\mathbb{c}(i)}^p = 0.2$, $k_{\mathbb{c}(i)} = b_{\mathbb{c}(i)} = 100$, for the second class $t_{\mathbb{c}(i)}^z = t_{\mathbb{c}(i)}^p = 2$, $k_{\mathbb{c}(i)} = b_{\mathbb{c}(i)} = 120$, and for the third class $t_{\mathbb{c}(i)}^z = t_{\mathbb{c}(i)}^p = 20$, $k_{\mathbb{c}(i)} = b_{\mathbb{c}(i)} = 150$.
Such values indicate that the users wish to pay a unit of monetary value for a unit of offered resources. 
\par

\begin{figure}[t]
\begin{subfigure}{0.49\linewidth}
\centering
\includegraphics[height=4.5cm,width=1.0\columnwidth]{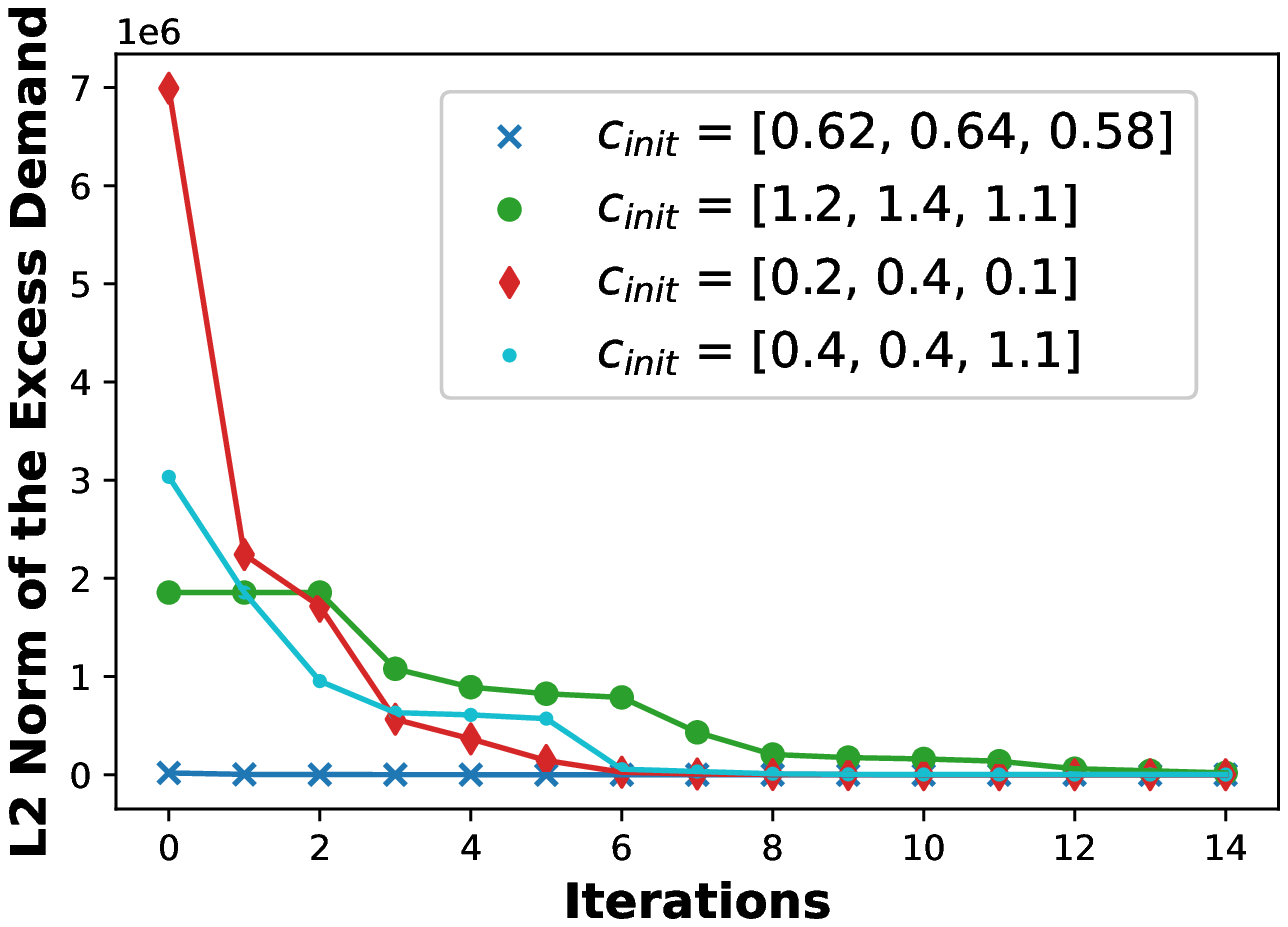}
\vspace{-17pt}
\caption{}
\label{fig:1a}
\end{subfigure}
\hspace*{\fill} 
\begin{subfigure}{0.49\linewidth}
\centering
\includegraphics[height=4.5cm,width=1.0\columnwidth]{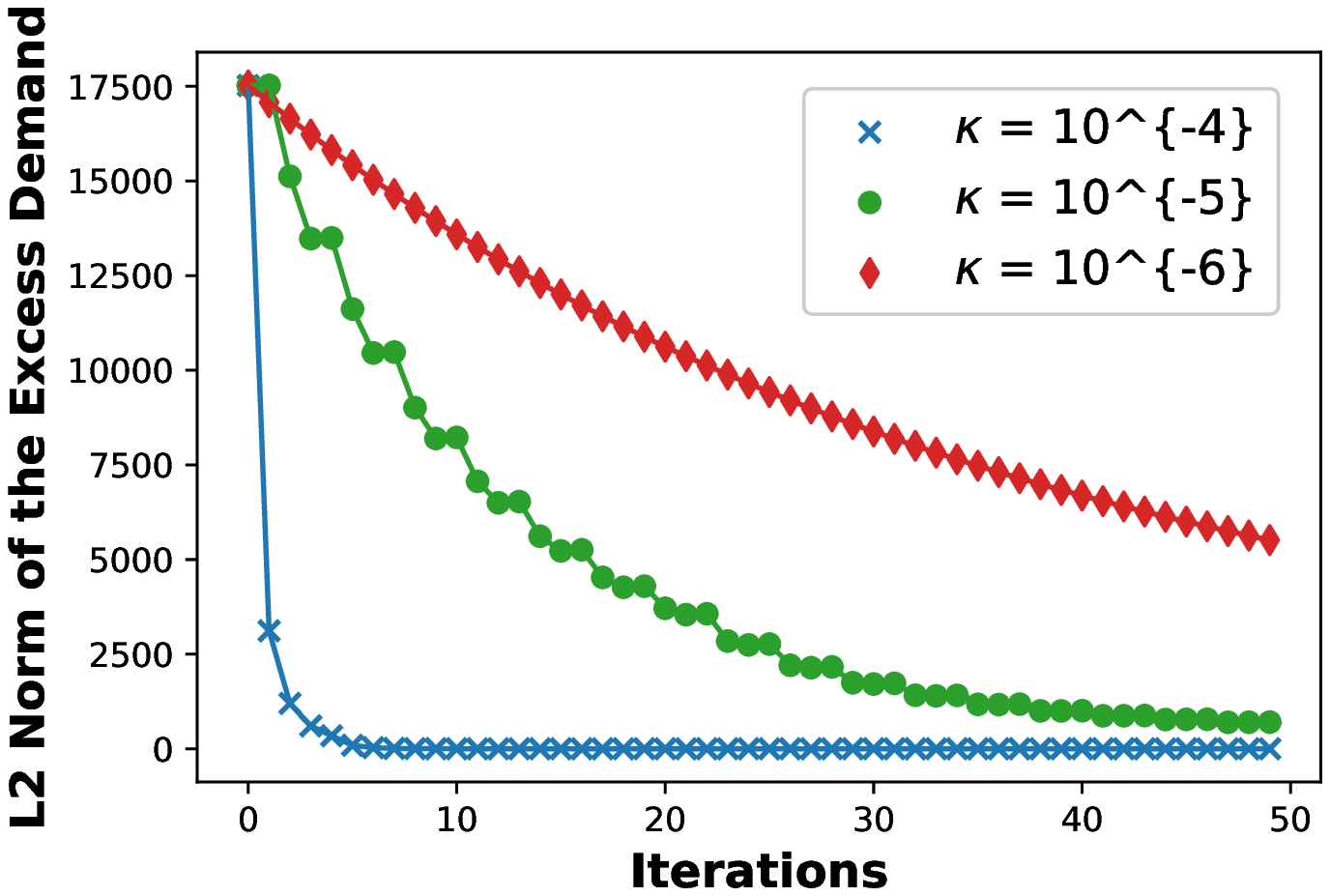}
\vspace{-17pt}
\caption{} 
\label{fig:1b}
\end{subfigure}
\vspace{-9pt}
\caption{\small{L2 norm of the excess demand vector throughout the clock auction (a) for $\kappa = 10^{-4}$ and various initialization price vectors $\bm{c}_{init}$, and (b) for $\bm{c}_{init}^T = [0.62, 0.64, 0.58]$ and different values of $\kappa$.}} 
\label{fig:1}
\end{figure}

\begin{figure}[t]
\centering
\includegraphics[width=0.6\columnwidth]{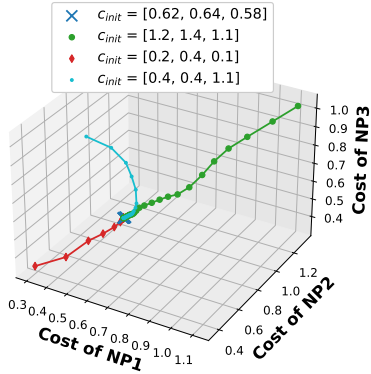}
\vspace{-5pt}
\caption{\small{Illustrating Theorem \ref{th:convergence}. Starting from any price vector $\bm{c}_{init}$, the clock auction converges to the market clearing prices $\bm{c^\dagger}$.}}
\label{fig:2}
\end{figure}

To discretize the auction, we change the cost vector according to a step value, $\kappa$, as $\bm{c}_{t+1} = \bm{c}_t + \kappa \mathcal{Z}(\bm{c}_t).$ Fig. \ref{fig:1} depicts the L2 norm of the excess demand vector throughout the clock auction for different cost vector initializations $\bm{c}_{init}$ (Fig \ref{fig:1a}), and for different step values $\kappa$ (Fig. \ref{fig:1b}). By simulating the clock auction, we deduce that the clearing price vector is $\bm{c}^{\dagger T} = [0.6116, 0.6273, 0.5811]$. In Fig. \ref{fig:1a} note that the closer the initialization cost vector is to $\bm{c}^{\dagger T}$, the faster the convergence becomes. 
Fig. \ref{fig:1b}, connotes the need for a proper choice of the step value $\kappa$. 
Clearly, $\kappa = 10^{-4}$ gives the fastest convergence and as we decrease the step values it becomes slower. Nevertheless, since Theorem \ref{th:convergence} is proved for the continuous case, large values of $\kappa$ cannot guarantee the convergence of the auction to an equilibrium.
In Fig. \ref{fig:2} observe that the convergence of the auction does not depend on the initialization of the cost vector
(Theorem \ref{th:convergence}).


\subsection{Visualization of the Resource Allocation}
In this section, we get insights on the allocation of the resources in the market. We assume $2$ NPs with $C_1 = C_2 = 1400$ and $2$ SPs with $10$ users each and one shared service class with $t_{\mathbb{c}(i)}^z = t_{\mathbb{c}(i)}^p = 0.2$ and $k_{\mathbb{c}(i)} = b_{\mathbb{c}(i)} = 100$ for all $i$. The first SP (SP1) is near the first NP (NP1) and far from NP2 and hence, we set $[\beta_{(1,1)},\dots, \beta_{(1,10)}] = [0. 99, 0.96, 0.87, 0.85, 0.82, 0.81, 0.80, 0.80, 0.70, 0.70]$ and $\beta_{(2,i)} = 0.2, \forall i \in \mathcal{U}_{1}$.
Moreover, for the users of SP2 we set $\beta_{1,i} = \beta_{2,i} = 0.8, \forall i \in \mathcal{U}_{2}$. 

\par 
We compare the resource allocation of $four$ different methods. First, 'Auction' refers to the resource allocation that results immediately 
after the auction. 'SPP' takes $\bm{\hat{x}}_m^{*}$ from the equilibrium but performs the intra-slice of every SP by solving $\bm{IN-SL}$. We also study the method 'oSPP(5\%)', which mimics the SPP method but with 5\% overbooked resources. Finally, 'SWM' refers to the solution of the Problem $\bm{SWM}$.

\begin{figure}[t]
\centering
\includegraphics[width=0.62\columnwidth]{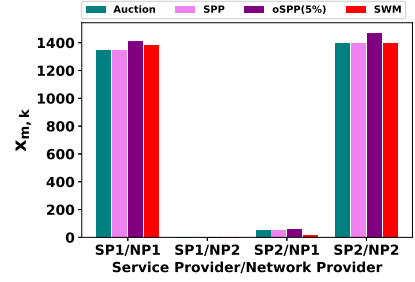}
\vspace{-5pt}
\caption{\small{Total amount of resources obtained by every SP $m$ from every  NP $k$ in the market, $\bm{x}_{(m,k)}$.}}
\label{fig:4}
\end{figure}

\par
Fig. \ref{fig:4} shows the amount of resources obtained from the two SPs. All methods allocate the majority of the resources of NP1 to SP1 since its users have greater connectivity with it. Although the users of SP2 have equally high connectivity with both NPs, all of the four methods were flexible enough to allocate the resources of NP2 to SP2. Note that none of the methods gives resources from NP2 to SP1. 

\begin{figure*}[!t]
\vspace{-\baselineskip}
\centering
\subfloat[]{
\includegraphics[height=4cm,width=5.5cm]{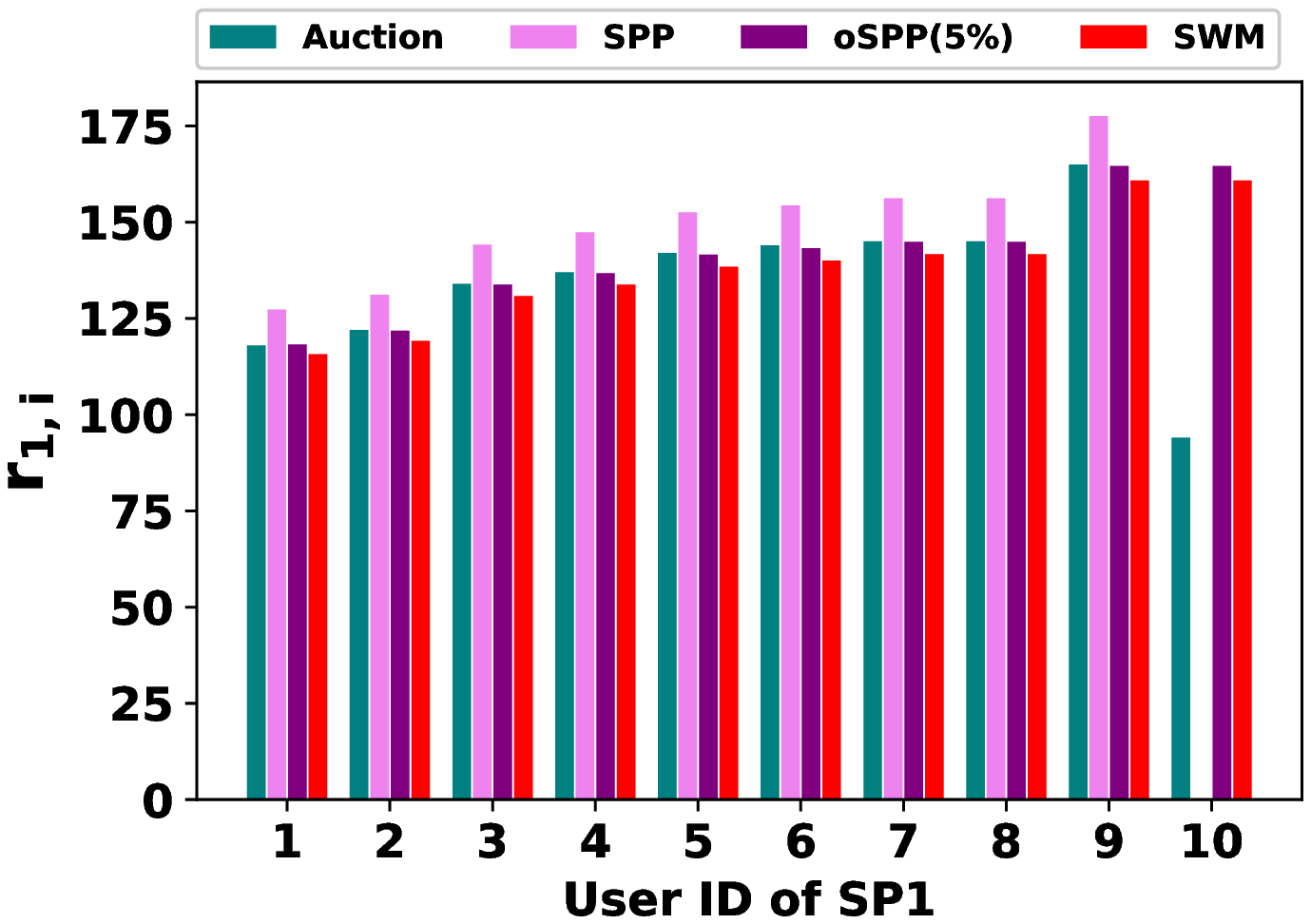}
\label{fig:5a}
\vspace{-5pt}}
\hspace*{-8pt}
\subfloat[]{
\includegraphics[height=4cm,width=5.5cm]{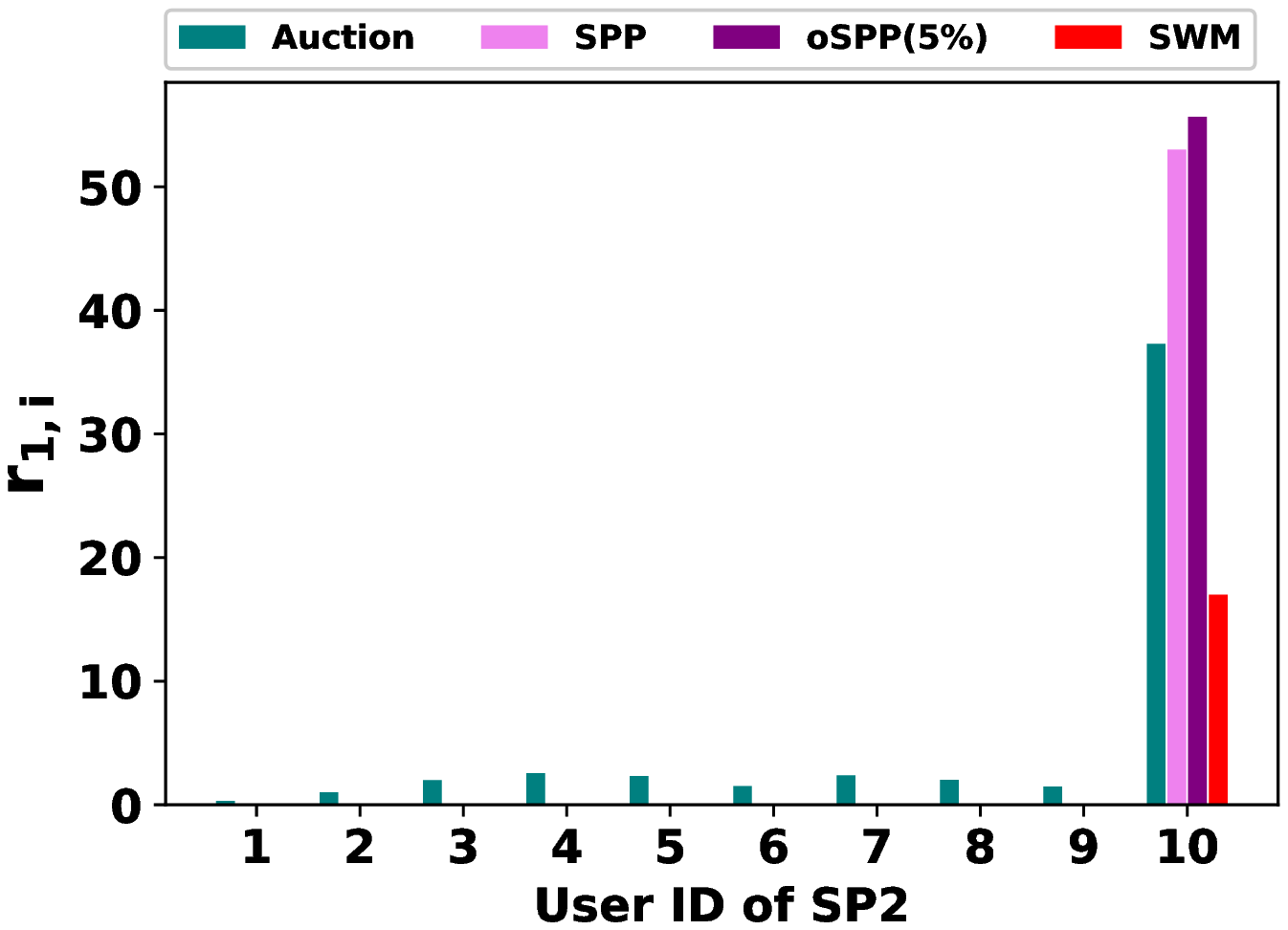}
\label{fig:5b}
\vspace{-5pt}}
\hspace*{-8pt}
\subfloat[]{
\includegraphics[height=4cm,width=5.5cm]{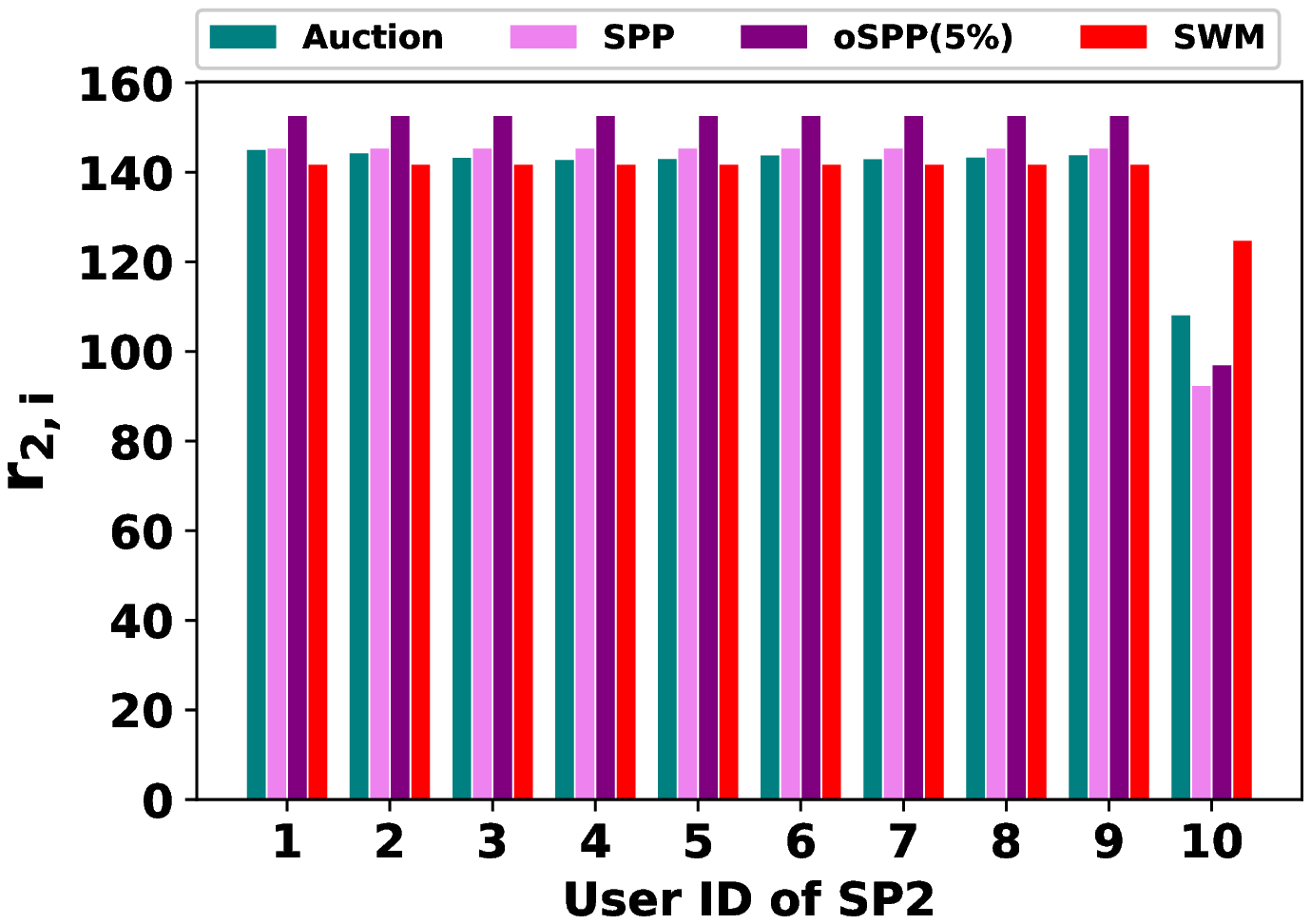}
\label{fig:5c}
\vspace{-5pt}}
\vspace{-5pt}
\caption{\small{The solution of the intra-slice resource allocation problem from the perspective of the two different SPs of the market. Specifically, how (a) SP1 distributed the resources of NP1, i.e., $\bm{r}_{1,i}$ for every i in $\mathcal{U}_1$, (b) SP2 distributed the resources of NP1, i.e., $\bm{r}_{1,i}$ for every i in $\mathcal{U}_2$, and (c) SP2 distributed the resources of NP2, i.e., $\bm{r}_{2,i}$ for every i in $\mathcal{U}_2$.}}
\label{fig:5}
\vspace{-1.0\baselineskip}
\end{figure*}

Fig. \ref{fig:5} depicts the intra-slice resource allocations.
In Fig. \ref{fig:5a} observe that the greater the connectivity of a user is, the less resources it gets.
That is because users with good connectivity factors meet their prerequisite QoS using less resources and hence SP1 could maximize its expected profit by giving them less. Note that 'SPP' gives no resources to the user with the worst connectivity whereas with the overbooking, SP1 gets enough resources to make attractive offers to every user. Therefore, 'SPP' might make an unfair allocation, since when the resources are not enough, it neglects the users with bad connectivity. 
In Fig. \ref{fig:5c}, note that the homogeneity in the connectivities of the users of SP2 forces every method to fairly divide the resources among them.

\par
\begin{figure*}[!t]
\centering
\subfloat[]{
\includegraphics[height=4cm,width=5.5cm]{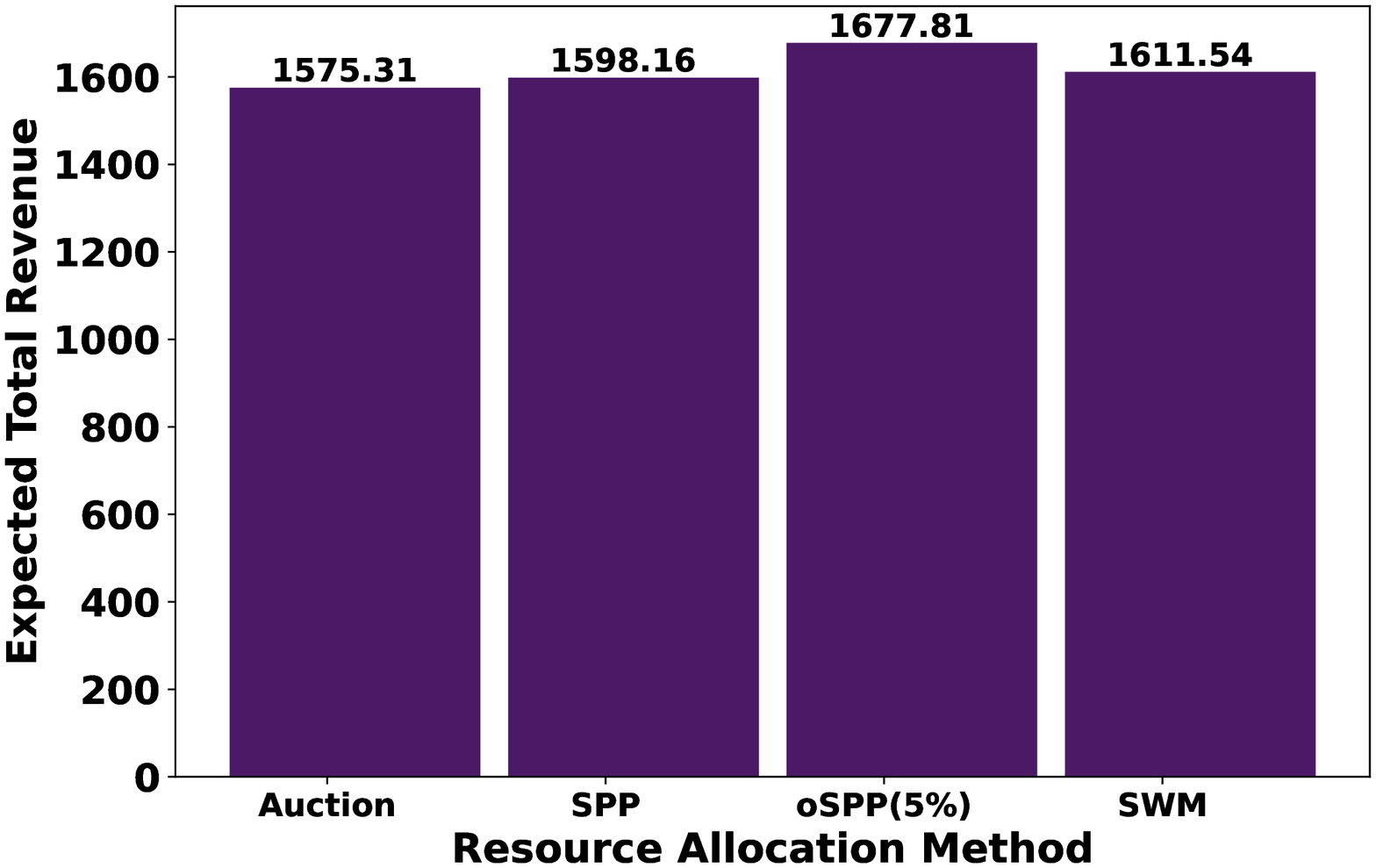}
\label{fig:6a}
\vspace{-5pt}}
\hspace*{-8pt}
\subfloat[]{
\includegraphics[height=4cm,width=5.5cm]{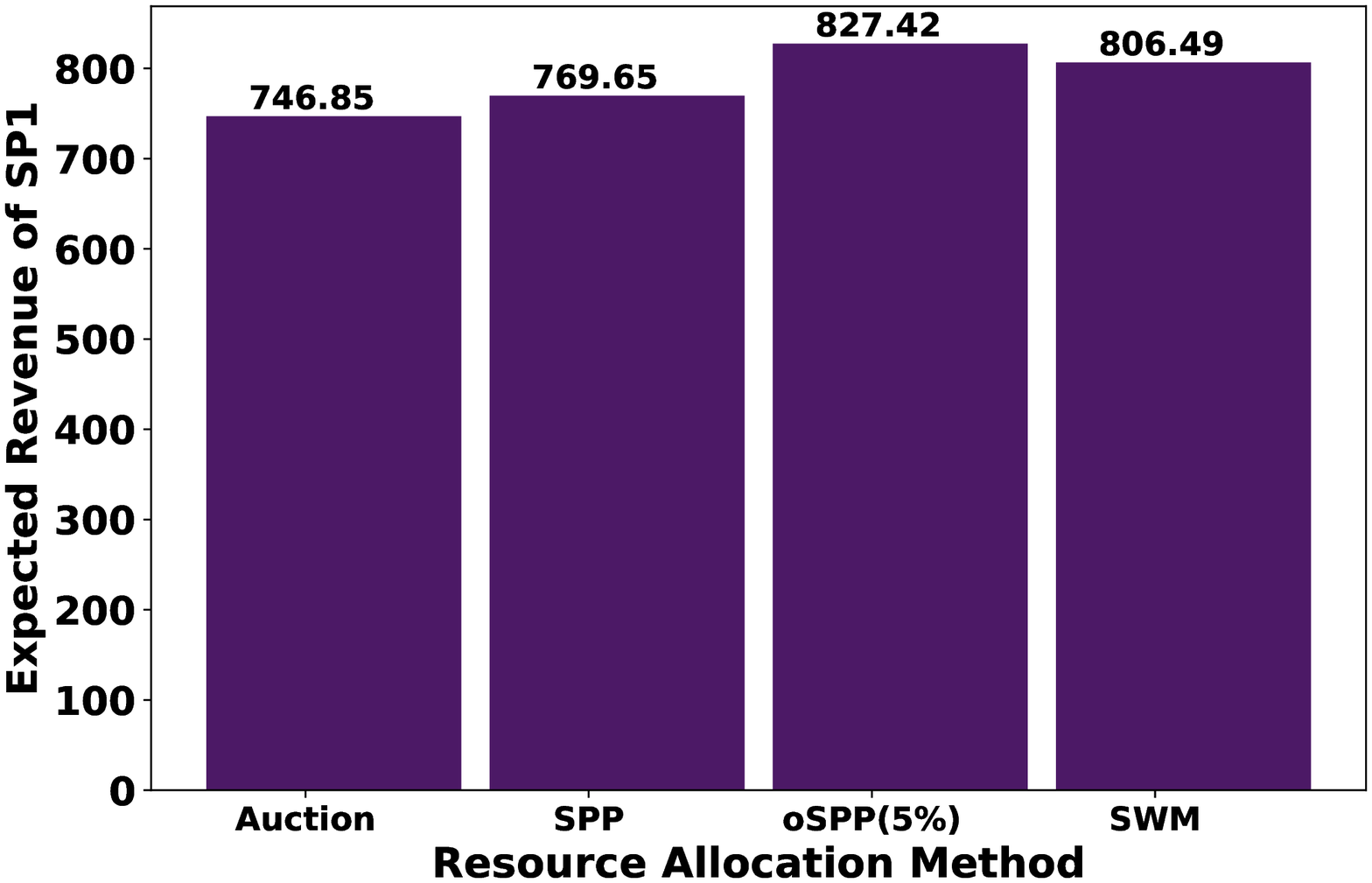}
\label{fig:6b}
\vspace{-5pt}}
\hspace*{-8pt}
\subfloat[]{
\includegraphics[height=4cm,width=5.5cm]{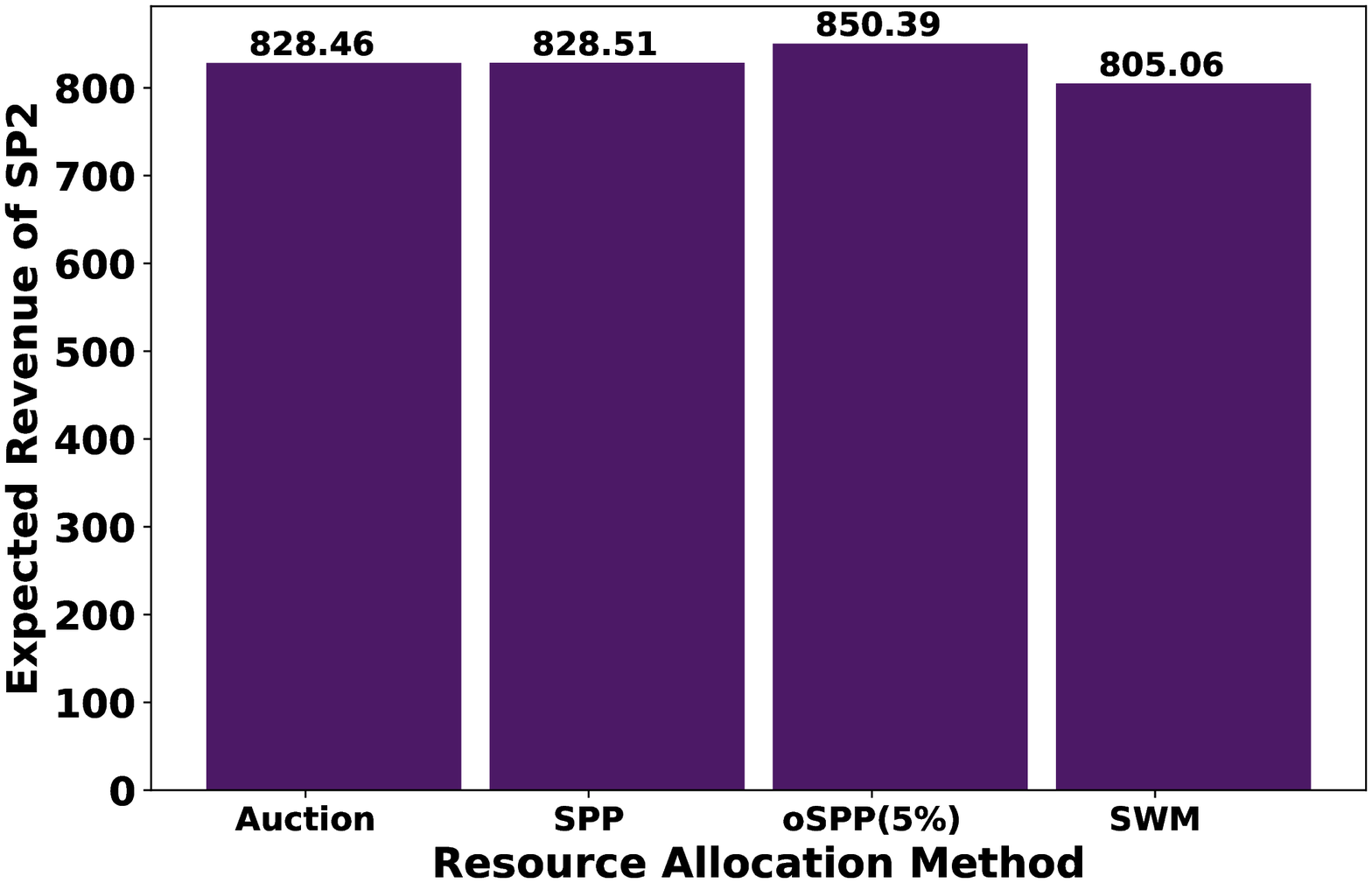}
\label{fig:6c}
\vspace{-5pt}}
\vspace{-5pt}
\caption{\small{Illustrating the expected revenue (given by Eq. \eqref{tt2}) for the four different resource allocation methods. Fig. (a) shows the aggregated expected revenue, Fig. (b) shows the expected revenue of SP1, and Fig. (c) shows the expected revenue of SP2.  }}
\label{fig:6}
\vspace{-1.0\baselineskip}
\end{figure*}

Fig. \ref{fig:6a} shows the expected value of the total revenue, or the social welfare. 'SWM' gives the greatest revenue among the methods that do not overbook. Nevertheless, although 'SPP' is a completely distributed solution and was not designed to maximize the total revenue, it performs very close to 'SWM'. Moreover, a $5\%$ overbooking leads to greater revenues. 


\subsection{Impact of Bayesian Inference}
The previous results are extracted after a sufficient number of cycles, when the SPs have learned the parameters of the end-users. In this section, we consider an SP with $10$ users and one service class that employs Bayesian inference to learn the private parameter $t_{\mathbb{c}(i)}^z$ for every $i$. We set the true value of the parameter to be $t_{\mathbb{c}(i)}^z = 2$. The other parameters are set $t_{\mathbb{c}(i)}^p = 2$, $k_{\mathbb{c}(i)}=b_{\mathbb{c}(i)} = 120$ and $\beta_{1,i} = 0.9, \forall i \in \mathcal{U}_1$. We assume one more SP with a unique service class with $t_{\mathbb{c}(i)}^p = t_{\mathbb{c}(i)}^z = 0.2$, $k_{\mathbb{c}(i)}=b_{\mathbb{c}(i)} = 100$ and $\beta_{2,i} = 0.9  \forall i \in \mathcal{U}_2$. Finally, there are 2 NPs with $C_1 = C_2 = 1200$. 
\par
\begin{figure}[t]
\begin{subfigure}{0.49\linewidth}
\centering
\includegraphics[width=0.97\columnwidth]{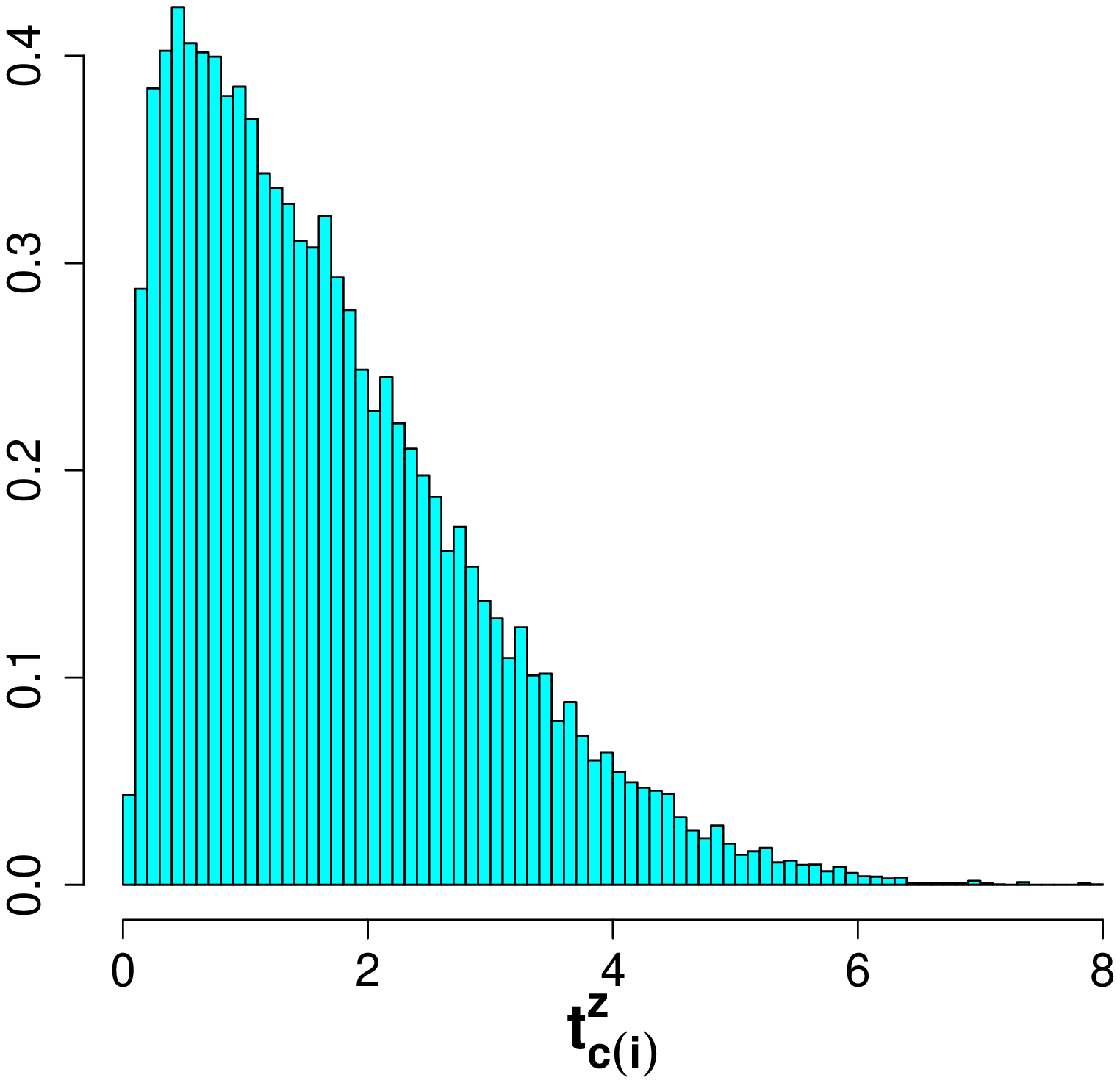}
\vspace{-7pt}
\caption{}
\label{fig:7a}
\end{subfigure}
\hspace*{\fill} 
\begin{subfigure}{0.49\linewidth}
\centering
\includegraphics[width=0.97\columnwidth]{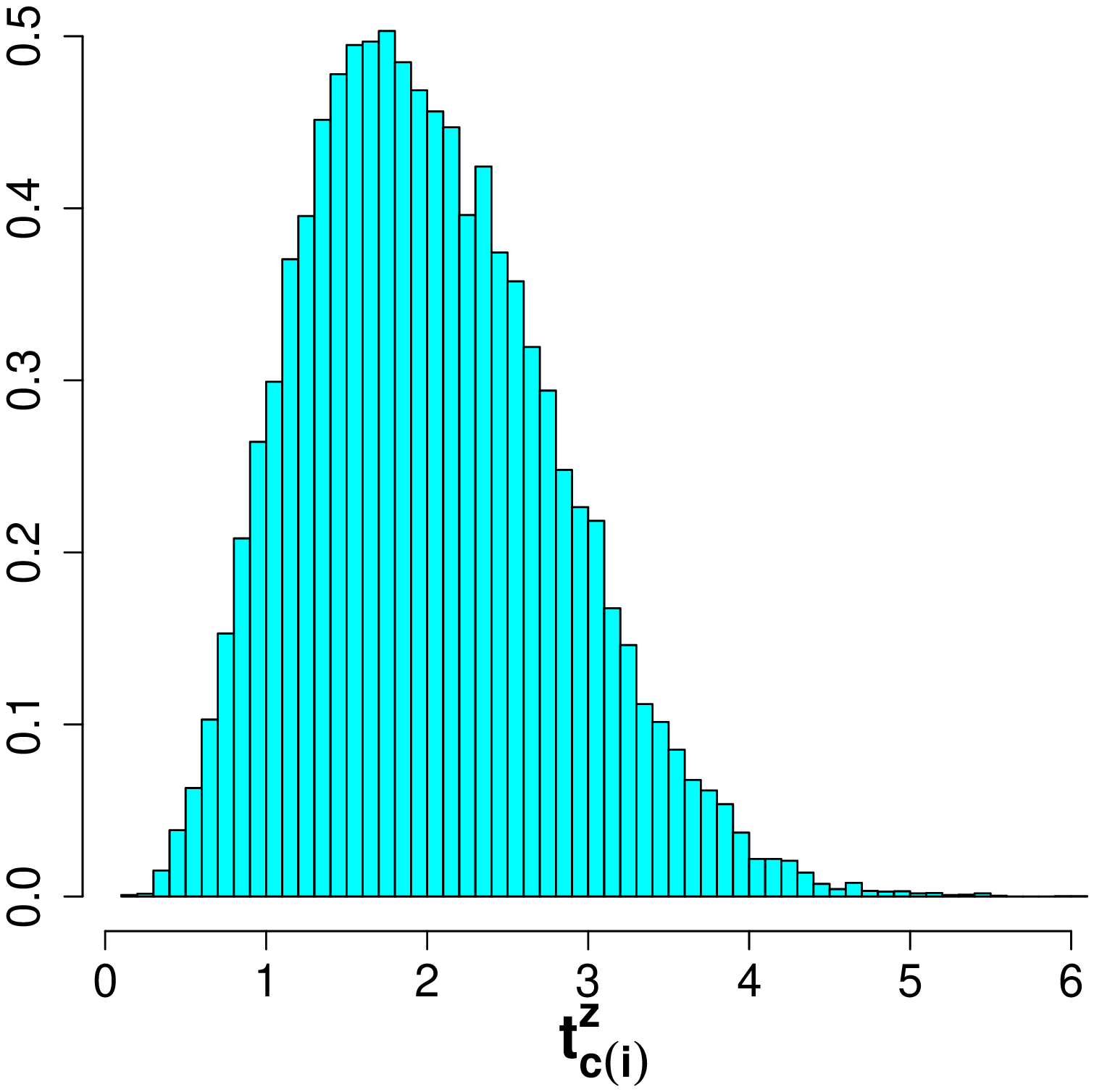}
\vspace{-7pt}
\caption{} 
\label{fig:7b}
\end{subfigure}
\vspace{-9pt}
\caption{\small{Posterior distribution of the unknown private parameter $t_{\mathbb{c}(i)}^z$ in (a) the first Market Cycle, and (b) in the second Market Cycle.}} 
\label{fig:7}
\end{figure}

In this example, SP1 sets as prior distribution the normal $\mathcal{N}(0.02, 2)$ and hence assumes elastic traffic. 
At the end of each market cycle, the SP makes an estimation, $\hat{t}_{\mathbb{c}(i)}^z$, by calculating the mean of the posterior distribution.
Fig. \ref{fig:7} depicts the histogram of the posterior distribution for the first two market cycles. Observe that even in the third market cycle, SP1 can estimate with high accuracy the actual value of the parameter. In Table \ref{table:1}, note that the perceived revenue, i.e., the expected revenue calculated using the estimation, is different between the cycles that $\hat{t}_{\mathbb{c}(i)}^z$ differs from ${t}_{\mathbb{c}(i)}^z$. Hence, it is impossible for the SPs to maximize their expected profits when they don't know the actual values of the parameters. Indeed, observe that the bad estimate of $\hat{t}_{\mathbb{c}(i)}^z = 0.02$ gives poor expected revenue compared to the last two cycles. 

\begin{table}[h!]
\centering
\begin{tabular}{|p{1cm}|p{1cm}|p{1.5cm}|p{1.5cm}|p{1.5cm}|  }
 \hline
 \vspace{0.1cm} Cycle\vspace{0.1cm}  & \vspace{0.1cm} $\hat{t}_{\mathbb{c}(i)}^z$ \vspace{0.1cm} & \vspace{0.1cm}Acquired Resources\vspace{0.1cm} &\vspace{0.1cm}Perceived Revenue\vspace{0.1cm} & \vspace{0.1cm}Actual Revenue\vspace{0.1cm}\\
 \hline
    &     & &  &   \\
  1   & 0.02    &1087 & 530.26 &   699 \\
 2  &  1.68  &1370 & 1160.77   &1163.48\\
 3  & 2.01 &1365 & 1161.42 &  1161.42\\

 \hline
\end{tabular}
\caption{Bayesian inference in different market cycles.}
\label{table:1}
\end{table}

\section{Concluding Remarks}

In this paper we focus on the technical and economic challenges that emerge from the application of the network slicing architecture to real world scenarios. Taking into consideration the heterogenity of the users' service classes we introduce an iterative market model along with a clock auction that converges to a robust  $\epsilon$-competitive equilibrium. Finally, we propose a Bayesian inference model, for the SPs to learn the private parameters of their users and make the next equilibria more efficient. Numerical results validate the convergence of the clock auction and the capability of the proposed framework to capture the different incentives. 

\clearpage

\bibliographystyle{IEEEtran}
\bibliography{IEEEabrv,bibliography}

\vspace{12pt}

\end{document}